\documentclass[prd,superscriptaddress,amsfonts,amssymb,amsmath,showpacs,twocolumn,nofootinbib]{revtex4-2}
\usepackage{bm}
\usepackage{amsfonts}
\usepackage{latexsym}
\usepackage{graphicx}
\usepackage{amsmath}
\usepackage{palatino}
\usepackage{xcolor} 
\usepackage{mathpazo}
\usepackage{xcolor}
\usepackage{rotating}
\usepackage{adjustbox}
\usepackage{tensor}
\usepackage{textcomp}
\usepackage{float}
\usepackage{booktabs}
\usepackage{dcolumn}
\usepackage[titletoc]{appendix}
\usepackage{booktabs}
\usepackage{multirow}
\usepackage{hyperref}
\hypersetup{colorlinks,citecolor=blue}
\usepackage{amsmath}
\usepackage{xcolor}
\usepackage{orcidlink}
\usepackage{epsfig}
\usepackage{caption}
\usepackage{subcaption}
\usepackage{commath}
\hypersetup{colorlinks,citecolor=blue}
\hypersetup{colorlinks=true,linkcolor=magenta,filecolor=magenta,    urlcolor=blue}

\def\be{\begin{equation}}
\def\ee{\end{equation}}
\def\bea{\begin{eqnarray}}
\def\eea{\end{eqnarray}}

\begin{document}

\title{Evidence for evolving dark energy from DESI DR2 BAO and Pantheon$^+$, DES-Dovekie, and Union3}

\author{Himanshu Chaudhary}
\email{himanshu.chaudhary@ubbcluj.ro,\\
himanshuch1729@gmail.com}
\affiliation{Department of Physics, Babeș-Bolyai University, Kogălniceanu Street, Cluj-Napoca, 400084, Romania}
\affiliation{Research Center of Astrophysics and Cosmology, Khazar University, Baku, AZ1096, 41 Mehseti Street, Azerbaijan}
\author{Salvatore Capozziello}
\email{capozziello@na.infn.it}
\affiliation{Dipartimento di Fisica ``E. Pancini", Universit\`a di Napoli ``Federico II", Complesso Universitario di Monte Sant’ Angelo, Edificio G, Via Cinthia, I-80126, Napoli, Italy,}
\affiliation{Istituto Nazionale di Fisica Nucleare (INFN), sez. di Napoli, Via Cinthia 9, I-80126 Napoli, Italy,}
\affiliation{Scuola Superiore Meridionale, Via Mezzocannone 4, I-80134, Napoli, Italy.}
\author{Vipin Kumar Sharma}
\email{vipin.339122@lpu.co.in.}
\affiliation{International Center for High Energy Physics and Applications, Lovely Professional University, Phagwara, Punjab, 144411, India}
\author{Isidro Gómez-Vargas}
\email{isidro.gomezvargas@unige.ch}
\affiliation{Department of Astronomy, University of Geneva, Chemin Pegasi 51, Versoix, Geneva, 1290, Switzerland.}
\author{G. Mustafa}
\email{gmustafa3828@gmail.com}
\affiliation{Department of Physics,
Zhejiang Normal University, Jinhua 321004, People’s Republic of China}

\begin{abstract}
Evidences for evolving dark energy are shown using baryon acoustic oscillation measurements from the recent Dark Energy Spectroscopic Instrument Data Release 2 , combined with different Type Ia supernova datasets (Pantheon$^+$, DES-Dovekie, and Union3) and the CMB compressed likelihood. We examine several dark energy parameterizations, including the Logarithmic, Exponential, CPL, BA, JBP, Thawing, Mirage, and GEDE models. Analyzing the DESI DR2 measurements alone, we find that evidence for evolving dark energy is primarily driven by the LRG1-2 tracers, as their inclusion yields a preferred value of $w_0 > -1$. However, as each tracer provides only limited observables, this preference can result in an underconstrained and potentially unstable inference. Further, we find that each dark energy model predicts values in the $w_0 > -1$, $w_a < 0$ quadrant, a region characterized by the Quintom-B type dark energy scenario. The logarithmic bayes factor shows that, among all models, the Mirage model shows the inconclusive-to-moderate evidence across all dataset combinations. Consistently, the statistical significance remains modest, with $N\sigma \sim 1.1$-$2.3$, and no model showing a robust preference for dynamical dark energy using late-time datasets alone. The evolution of $w(z)$ shows a phantom crossing around $z \sim 0.5$ in most dynamical dark energy models, and the evolution of $f_{\mathrm{DE}}(z)$ converges to $f_{\mathrm{DE}}(0) = 1$ in all dark energy models.
\end{abstract}

\maketitle

\section{Introduction}\label{sec_1}
Dark energy (DE) has been postulated to account for the observed accelerated expansion of the Universe \cite{supernovaSearchTeam:1998fmf,supernovaCosmologyProject:1998vns}. The advent of a new era of precision cosmology has been made possible by recent developments in large-scale structure surveys, which have allowed for rigorous testing of the conventional positive cosmological constant with cold dark matter, i.e., $\Lambda$CDM paradigm \cite{Bamba:2012cp}. Among these, the Dark Energy Spectroscopic Instrument (DESI) is notable for its use of Baryon acoustic oscillations (BAO) observations to track cosmic expansion \cite{Sousa-Neto:2025gpj,adame2025desi,DESI:2024aqx,karim2025desi,lodha2025desi,lodha2025extended}. By analyzing standard cosmological rulers such as BAO, it is possible to place constraints on whether DE operates as a cosmological constant or exhibits time-dependent dynamical behavior \cite{Notari:2024zmi,dymnikova1998self,dymnikova2000decay,dymnikova2001decay,ray2011phenomenology,doroshkevich1984formation}.

Planck’s first-year data release in 2013~\cite{ade2014planck} indicated $w = -1.13^{+0.13}_{-0.14}$, slightly favoring the phantom regime. Subsequent improvements in supernova calibration through the 2014 Joint Light-curve Analysis (JLA) dataset~\cite{betoule2014improved} alleviated this tension, and combining JLA with Planck2013 results brought dark energy constraints into agreement with $\Lambda$CDM. The Planck-15 analysis~\cite{ade2016planck,ade2016planck}, which adopted JLA as its default supernova dataset, confirmed this consistency, reporting $w = -1.006^{+0.085}_{-0.091}$. More recently, the 2022 Pantheon$^+$ compilation~\cite{brout2022pantheon} found $w = -0.90 \pm 0.14$ from supernova data alone, and $w = -0.978^{+0.024}_{-0.031}$ when combined with CMB and BAO measurements. While these results remain consistent with $\Lambda$CDM within $2\sigma$, they show a mild deviation relative to earlier datasets. This trend is reinforced by the Union3 compilation~\cite{rubin2025union}, which reports a mild $1.7$–$2.6\sigma$ tension with $\Lambda$CDM, favoring models where $\omega_0 > -1$ and $\omega_a < 0$. Collectively, these observations point toward a potentially evolving dark energy component, with an equation of state that increases over time and a present value $w > -1$.

In 2024, building on hints of the evolving nature of dark energy from the Pantheon$^+$ and Union3 compilations, DES-SN5YR found that whether using supernova data alone or in joint fits with the CMB, BAO, and three two point (3$\times$2pt) measurements which refer to the joint analysis of three two-point correlation functions: galaxy clustering, galaxy–galaxy lensing, and cosmic shear, the best fit equation of state (EoS $w$ is consistently slightly greater than $-1$ at more than the $1\sigma$ level. This behavior agrees with results from Union3 and supports a trend toward mildly dynamical dark energy. In the same year, DESI released its first year BAO data \cite{adame2025desi}; When combined with CMB, Pantheon$^+$, Union3, and DES-SN5YR, these measurements yield $(w_0,w_a)$ constraints that depart from $\Lambda$CDM at the levels $2.6\sigma$, $2.5\sigma$, $3.5\sigma$, and $3.9\sigma$, respectively. DESI DR2 BAO measurements \cite{karim2025desi}, when combined with data from CMB only, exclude $\Lambda$CDM with a significance of $3.1\sigma$. Including additional supernova data sets Pantheon$^+$, Union3, and DES SN5YR yields exclusion significances of $2.8\sigma$, $3.8\sigma$, and $4.2\sigma$, respectively. Recently, the DES Collaboration reported improved cosmological constraints from a re-analysis of the DES-SN5YR catalog, showing that the preference for dynamical dark energy is reduced to a statistical significance of $3.2\sigma$, compared to the earlier $4.2\sigma$ result obtained with DES-SN5YR \citep{popovic2025dark}. Yet several studies have also shown that, in the DES-SN5YR dataset, the evidence for dynamical dark energy is biased by low-redshift SNe~Ia \cite{huang2025desi,gialamas2025interpreting,efstathiou2025evolving,cortes2025desi}. Compared to DR1, DR2 offers tighter constraints with improved precision and reduced uncertainties. Following the DR2 BAO release, \cite{lodha2025desi,lodha2025extended} performed an extended analysis of dark energy dynamics and confirmed evidence of a time-varying EoS. Evidence for a time-evolving dark energy model with a statistical significance exceeding $5\sigma$ has also been reported \citep{scherer2025challenging}. 

The high-precision BAO measurements from the DESI LRG1 sample, which span redshifts of approximately $z\approx 0.4$–$0.6$, probe epochs within the DE dominated era. As such, these LRG datasets are particularly sensitive to the signatures of cosmic acceleration, making them valuable for testing time-dependent DE models \citep{BARUA2025101995,Vilardi:2024cwq}. Numerous studies have reported hints of dynamical dark energy using DESI measurements, supporting the possibility of deviations from the $\Lambda$CDM model \cite{yin2024cosmic,cortes2024interpreting,carloni2025does,roy2025dynamical,orchard2024probing,giare2024robust,pang2025constraints,wolf2025robustness,gao2024evidence,giare2025overview,colgain2025much,park2024w_0w_a,fazzari2025cosmographic,moffat2025dynamical,mirpoorian2025dynamical,toomey2025theory,ishak2025fall,qiang2025new,capozziello2025dark,sharma2025probing,chaudhary2025lambdacdm,capozziello2025evidence,chaudhary2025does,chaudhary2025probing}. In parallel, model-independent and non-parametric approaches have been developed to reconstruct the expansion history without assuming a specific form for $w(z)$ \cite{mukherjee2024model,dinda2025model,jiang2024nonparametric,tsedrik2025interacting,li2025reconstructing,wang2025model}. Alongside these efforts, alternative theoretical frameworks including holographic dark energy, ricci dark energy, non-cold dark matter, exotic dark matter scenarios, braneworld models and monodromic dark energy have also been explored in light of the DESI measurements \cite{wu2025observational,plaza2025probing,li2025exploring,braglia2025exotic,mishra2025braneworld,goldstein2025monodromic}. In particular, quintessence-based interpretations of evolving dark energy in light of DESI DR2 have been widely explored, with scalar-field dynamics \cite{tada2024quintessential,bhattacharya2024cosmological,ramadan2024desi,gialamas2025quintessence,dinda2025physical,adam2025comparing,petri2025dark,wang2025resolving,berghaus2024quantifying}. In addition, modified theories of gravity have been investigated using the DESI measurements \cite{mazumdar2025constraint,liu2025torsion}, while potential tensions between DESI DR2 and CMB observations have also been examined \cite{ye2025tension}. The role of topological effects in shaping the behavior of dynamical dark energy has likewise been explored \cite{an2025topological}. In a related direction, several studies have also investigated the impact of $\sum m_{\nu}$ in the context of DESI DR1/DR2 using extended twelve-parameter dynamical dark energy models and late-time dark energy scenarios. \cite{choudhury2024updated,choudhury2025cosmology,rebouccas2025investigating,silva2025new}. Indeed, these developments have also raised questions about the status of the $H_0$ tension in light of DESI measurements, which suggest evidence for dynamical dark energy and may point toward the need for new physics to resolve this discrepancy \cite{wang2024dark,vagnozzi2020new,vagnozzi2023seven,pedrotti2025bao}

Motivated by the above studies, in this paper we explore several dark energy models beyond the CPL parametrization, focusing exclusively on late-time datasets. In particular, we compare phenomenological constraints on the dark energy equation of state $w$ derived with and without DESI LRG1 data. We further extend this analysis by incorporating different SNe~Ia samples together with DESI DR2 and compressed CMB likelihoods. This approach allows us to assess the statistical significance of dynamical dark energy features and to isolate the role of late-time observations in driving the inferred evolution. Our paper is organized as follows. In Section \ref{sec_2}, we introduce the cosmological background equations and models. Section \ref{sec_3} details the core of this work with datasets and methodology using the Markov Chain Monte Carlo (MCMC) sampling against the publicly available DESI DR2 data, while section \ref{sec_4} is dedicated to the discussion of results.  In Section \ref{sec_5}, we draw the conclusions.

\section{Background Equations and Dark energy models}\label{sec_2}
Assuming a spatially flat Friedmann-Lema\^{i}tre-Robertson-Walker (FLRW) cosmological background 
\begin{eqnarray}
ds^{2}= -dt^{2} + a^2(t)[dr^2 + r^2 (d\theta^{2} + \sin^{2}\theta d\phi^{2})] \label{a1},\end{eqnarray}
where $a(t)$ is the time-dependent cosmological scale factor and ($r$,$\theta$,$\phi$) are the standard  spherical coordinates,
and considering the late-time Universe —specifically, where radiation can be neglected— the first Friedmann equation arising from the Einstein field equations
\begin{eqnarray}
G_{\mu\nu}=8\pi G\, {T^{(m)}_{\mu\nu}}\label{a4},\end{eqnarray} 
takes the form:
\begin{equation}
    H^2 = \frac{8\pi G}{3} \left(\rho_\textrm{m} a^{-3} +  \rho_{\textrm{de}} \right).
    \label{eqn:Hubble_1}
\end{equation}
 where $T^{(m)}_{\mu\nu}$ is the standard matter-energy component, $G_{\mu\nu} = R_{\mu\nu}-g_{\mu\nu} R/2$ is the Einstein tensor, and $G$ is the Newtonian gravitational constant.
The continuity equation in FLRW background reads
\begin{equation}\label{eq_2}
\dot{\rho}_{\text{x}} + 3H(1 + w_{\text{x}}) \rho_{\text{x}} = 0,
\end{equation}
where \(\rho_\text{x}\) represents the energy density of each component with x$\in$ (matter, DE), over $(\dot{})$ represents the cosmic time derivative, and $w_x$ represents the equation of state parameter (EoS). 

The expansion history of the Universe is governed by the dimensionless Hubble parameter ${\displaystyle E(z) = \frac{H(z)}{H_0}}$, where ${\displaystyle H(z) = \frac{\dot{a}}{a}}$ is the Hubble parameter in function of the redshift $z$, and $H_0$ is its present-day value i.e., at $z=0$. Using Eq. \eqref{eqn:Hubble_1}, the functional $E(z)^2$ is given by:

\begin{equation}\label{Ez}
E(z)^2 = \Omega_m(1 + z)^3 + \Omega_r(1 + z)^4 + \Omega_\Lambda f_{\mathrm{DE}}(z),
\end{equation}
where $\Omega_m$, $\Omega_r$, and $\Omega_\Lambda$ are the present-day fractional densities of matter, radiation, and dark energy, respectively. Also, here $f_{DE}$ represents the evolution of DE \cite{Benetti:2019gmo, Petreca:2023nhy}. 
The evolution of DE ($\rho_{\textrm{de}}$) within Eq.\eqref{eqn:Hubble_1} will be the following solution of Eq.\eqref{eq_2}:
\begin{equation}\label{eqn:rhode}
f_{\mathrm{DE}}(z) \equiv \frac{\rho_{\mathrm{DE}}(z)}{\rho_{\mathrm{DE},0}} = \exp\left[ 3 \int_0^z \frac{1 + w(z')}{1 + z'} \, dz' \right]
\end{equation}
where $\rho_{\rm de, 0}$ is the present value of the DE density. For the constant $w$, the \eqref{eqn:rhode} simply $(1+z)^{3(1+w)}$, and a cosmological constant corresponds to $w = -1$. We also consider the Mirage Dark Energy parametrization \cite{linder2007mirage}, constructed to reproduce the CMB distance to last scattering of $\Lambda$CDM while allowing a dynamical equation of state $w(a)=w_0 + w_a(1-a)$. It effectively yields $w \approx -1$ near the pivot scale factor $a \approx 0.7$. In this framework, the CPL parameters satisfy $w_a = -3.66 \left(1 + w_0 \right)$, and the Thawing Dark Energy parametrization, which satisfies $w_a = -1.58 \left(1 + w_0 \right).$ Apart from this, we also consider different dynamical dark energy models; see Table~\ref{tab_1}, where we present the corresponding EoS and $f_{\rm DE}(z)$ for those models.

Although no fundamental principle uniquely specifies the optimal functional form for parameterisations, observational data can be employed to constrain and select those formulations that are both cosmologically consistent and phenomenologically viable. In the following, we examine several distinct parameterisation schemes for describing the evolution of dark energy (DE).
\begin{table*}
\begin{tabular}{|c|c|c|c|}
\hline
\textbf{Parameterization} & \textbf{$w$$(z)$} & \textbf{$f_{DE}(z)$} & \textbf{Reference} \\
\hline

Logarithmic &
$w_0 + w_a \log(1+z)$ &
$(1+z)^{3(1+w_0)} e^{\frac{3}{2} w_a(\log(1+z))^2}$ &
\cite{efstathiou1999constraining,silva2012thermodynamics} \\

Exponential & $w_0 + w_a \left[ \frac{z}{1+z} + \frac{1}{2}\left(\frac{z}{1+z}\right)^2 \right].$ &
$e^{\!\left[3 w_a\!\left(\frac{-z}{1+z}\right)\right]\,
(1+z)^{3(1+w_0+w_a)}\,}
e^{\!\left[3 w_a\!\left(\frac{1}{4(1+z)^2}+\frac{1}{2(1+z)}-\frac{3}{4}\right)\right]\,
(1+z)^{\tfrac32 w_a}}$ &
\cite{najafi2024dynamical} \\

CPL &
$w_0 + \frac{z}{1+z} w_a$ &
$(1+z)^{3(1+w_0+w_a)} e^{-\frac{3 w_a z}{1+z}}$ &
\cite{chevallier2001accelerating,linder2003exploring} \\

BA &
$w_0 + \frac{z(1+z)}{1+z^2} w_a$ &
$(1+z)^{3(1+w_0)}(1+z^2)^{\frac{3 w_a}{2}}$ &
\cite{barboza2008parametric} \\

JBP &
$w_0 + \frac{z}{(1+z)^2} w_a$ &
$(1+z)^{3(1+w_0)} e^{\frac{3 w_a z^2}{2(1+z)^2}}$ &
\cite{jassal2005wmap} \\

GEDE &
$-1-\frac{\Delta}{3\ln (10)}\!\left[ 1+\tanh\!\left( \Delta \log_{10}\!\left(\frac{1+z}{1+z_t}\right)\right)\right]$ &
$\left( \frac{1 - \tanh\!\left(\Delta \log_{10} \left( \frac{1+z}{1+z_t} \right)\right)}%
{1 + \tanh\!\left(\Delta \log_{10} (1 + z_t)\right)} \right)$ &
\cite{li2020evidence} \\

\hline
\end{tabular}
\caption{Dark energy parameterizations with their equations of state $w$$(z)$ and evolution functions $f_{DE}(z)$.}
\label{tab_1}
\end{table*}

\section{Dataset and Methodology}\label{sec_3}
To constrain the free parameters of each DE model considered in this work, we use the cosmological inference code \texttt{SimpleMC}\footnote{\url{https://github.com/ja-vazquez/SimpleMC.git}}. In our analyses, we use dynamic nested sampling \cite{feroz2008multimodal,feroz2013importance} using the \texttt{dynesty} Python package \cite{speagle2020dynesty}, as implemented within \texttt{SimpleMC}. To determine the number of live points, we follow the general rule of using $50 \times n_{\mathrm{dim}}$, where $n_{\mathrm{dim}}$ denotes the number of free parameters of the corresponding DE model.

We use the Bayesian evidence selection criterion to assess whether the DE models are statistically preferred over the $\Lambda$CDM model. The Bayesian evidence, $Z$, is defined as the integral of the likelihood over the prior volume, $Z = \int_{\Omega} \mathcal{L}(\theta)\,\pi(\theta)\, d\theta,$ where $\mathcal{L}(\theta) \equiv P(D|\theta, M)$ is the likelihood of the data $D$ given the parameters $\theta$ and model $M$, and $\pi(\theta) \equiv P(\theta|M)$ is the prior distribution of the parameters within that model. To compare two competing models, we compute the Bayes factor in logarithmic form, $\ln \mathcal{B}_{ij} = \ln\mathcal{Z}_i - \ln\mathcal{Z}_j,$ where $\ln\mathcal{Z}_i$ and $\ln\mathcal{Z}_j$ denote the Bayesian evidences of models $i$ and $j$, respectively. In our case, we report the Bayes factors $\ln \mathcal{B}_{ij}$ for the dynamical DE models relative to the $\Lambda$CDM model, based on the current observational data. Here, $i$ denotes the dynamical DE model, while $j$ denotes the $\Lambda$CDM model. Consequently, $\ln \mathcal{B}_{ij} > 0$ ($< 0$) indicates a preference for the DDE model ($\Lambda$CDM). The strength of model preference is interpreted using the Jeffreys scale \cite{kass1995bayes}: values of $\ln \mathcal{B}_{ij} < 1$ are considered inconclusive; $1 \leq \ln \mathcal{B}_{ij} < 2.5$ indicate weak evidence; $2.5 \leq \ln \mathcal{B}_{ij} < 5$ correspond to moderate evidence; $5 \leq \ln \mathcal{B}_{ij} < 10$ signify strong evidence; and $\ln \mathcal{B}_{ij} \geq 10$ is regarded as decisive evidence. In our analysis, the Bayesian evidence is computed by the \texttt{dynesty} package by algorithm.

The obtained results are subsequently analyzed and visualized using the \texttt{GetDist} package~\cite{lewis2025getdist}. Our analysis is based on data obtained from Baryon Acoustic Oscillation measurements, Type Ia supernovae, and Compressed CMB likelihood, which are detailed below:
\begin{itemize}
     \item \textbf{Baryon Acoustic Oscillation :} In our analysis, we use the recent BAO measurements from DESI Data Release 2 (DR2) \cite{karim2025desi}. These measurements are extracted using various tracers such as the Bright Galaxy Sample (BGS), Luminous Red Galaxies (LRG1–3), Emission Line Galaxies (ELG1–2), Quasars (QSO), and Lyman-$\alpha$ forests. To incorporate these measurements, one must compute the Hubble distance $D_H(z) = \frac{c}{H(z)}$, the comoving angular diameter distance $D_M(z) = c \int_0^z \frac{dz'}{H(z')}$, and the volume-averaged distance $D_V(z) = \left[ z, D_M^2(z), D_H(z) \right]^{1/3}$. It is necessary to derive the following ratios: $D_M/r_d$, $D_H/r_d$, $D_V/r_d$, and $D_M/D_H$ to constrain the parameters of each model, where $r_d$ is the sound horizon. In flat $\Lambda$CDM, it is $r_d = 147.09 \pm 0.2$ Mpc \cite{aghanim2020planck}.

     \item \textbf{Type Ia supernovae :} We also use three different SNe Ia compilations to improve the constraints on cosmological parameters. Among them, the Pantheon$^+$ (PP) sample \cite{brout2022pantheon} includes 1,701 light curves from 1,550 SNe Ia observations spanning the redshift range $0.01 \leq z \leq 2.26$. We exclude light curves at $z < 0.01$, as such low redshift data are affected by significant systematic uncertainties due to peculiar velocities. We also use the re-calibrated 1,820 photometric light curves collected over five years by the Dark Energy Survey Supernova Program (DES-Dovekie)~\cite{popovic2025dark}, which includes 1,623 DES-discovered Type~Ia supernovae and 197 externally sourced low-$z$ supernovae from the CfA and CSP samples~\cite{hicken2009cfa3,hicken2012cfa4,foley2017foundation}. The revised DES-Dovekie has 1,718 SNe~Ia overlapping between DES-Dovekie and DES-SN5YR~\citep{abbott2024dark}, and  Union3 compilation of 2087 cosmologically useful SNe~Ia from 24 datasets over a redshift range from \( 0.050 \leq z \leq 2.26 \) \cite{rubin2025union}. In our analysis, we marginalize over $\mathcal{M}$ parameter; for further details, see Equations (A9–A12) of \cite{goliath2001supernovae}.
     
     \item \textbf{Compressed CMB likelihood:} Finally, we use the compressed CMB likelihood because the dynamical dark energy models considered affect only the late-time expansion history of the Universe and mainly modify the geometrical features of the CMB. The full CMB spectrum includes small non-geometrical anomalies, such as the lensing amplitude and the low-$\ell$ power deficit, which may reflect residual systematics and bias dark energy estimates. For example, Planck data alone show a $\gtrsim 2\sigma$ preference for phantom dark energy~\citep{escamilla2024state}, mostly due to the lack of large-scale power. To avoid such biases, we rely on the compressed CMB likelihood. We use the CMB information compressed into the three parameters $\{\ \theta_\star, \omega_b,\, \omega_{cb}\,\}$, which is modeled as a $3\times3$ Gaussian likelihood~\citep{karim2025desi} (see Appendix~A, Eqs.~A1 and~A2).
\end{itemize}
In this analysis, the radiation density parameter is defined as  $\Omega_{r} = 2.469 \times 10^{-5}\, h^{-2} \left( 1 + 0.2271\, N_{\mathrm{eff}} \right)$~\cite{komatsu2009five}, where $N_{\mathrm{eff}} = 3.04$ represents the standard effective number of relativistic species.  The dark energy density parameter is determined from the flatness condition, $\Omega_{\Lambda} = 1 - \Omega_{r} - \Omega_{m}$, so that both $\Omega_{r}$ and $\Omega_{\Lambda}$ are computed from the remaining cosmological parameters. In this analysis, we use several DE models, with the chosen priors summarized in Table~\ref{tab_1}
\begin{table}[t] 
    \centering
    \begin{tabular}{|lll|}
    \hline
    Parametrization & Parameter & Prior\\  
    \hline 
    $\mathbf{\Lambda}$\textbf{CDM} & $H_0$ & $\mathcal{U}[40, 90]$ \\ 
    & $\Omega_{b}h^2$ & $\mathcal{U}[0.02, 0.025]$ \\
    & $\Omega_{m}$ & $\mathcal{U}[0.1, 0.5]$ \\
    \hline
    \textbf{Dark energy} & $w$ or $w_0$ & $\mathcal{U}[-3, 1]$ \\
    & $w_{a}$ & $\mathcal{U}[-3, 2]$ \\
    \hline 
    \textbf{Thawing/Mirage} & $w_0$ & $\mathcal{U}[-3, 1]$ \\ \hline
    \textbf{Emergent} & $\Delta$ & $\mathcal{U}[-3, 10]$ \\
    \hline
    \end{tabular}
\caption{The table shows the parameters and the priors used in our analysis for each DE model. The symbol $\mathcal{U}$ denotes that we use uniform priors}\label{tab_2}
\end{table}


\section{Results}\label{sec_4}
In this section, we explore various DE parameterizations to show the evidence for evolving DE. From this perspective, we first consider the DESI DR2 measurements alone to identify what drives the evolving nature of DE in this dataset. We then combine the DESI DR2 data with different Type~Ia supernova (SNe~Ia) measurements (Pantheon$^+$, DES-Dovekie, Union3) and a CMB compressed likelihood. Next, we compute the Bayes factor in logarithmic space $\ln \mathcal{B}_{i,j}$ to assess which model performs best among the models considered. Finally, we plot the evolution of the equation of state ($w$) and the normalized $(f_{\mathrm{DE}})$ as functions of redshift.

\subsection{Evidence for evolving dark energy in DESI BAO biased by LRG1 and LRG2 tracers}

\begin{figure}
\centering
\includegraphics[scale=0.55]{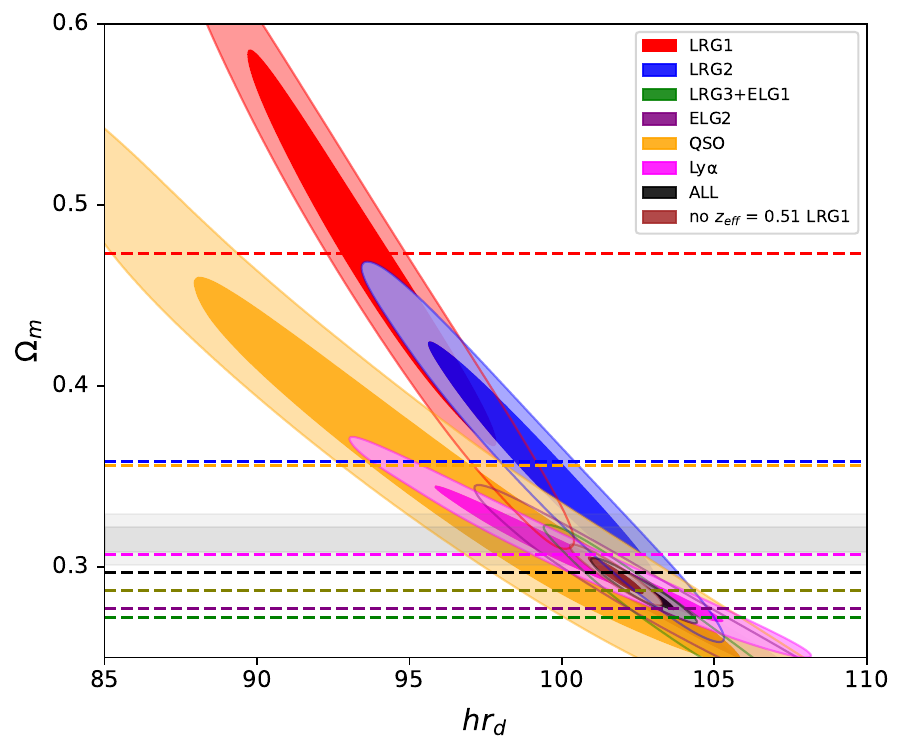}
\caption{The figure shows the posterior distributions at the 1$\sigma$ and 2$\sigma$ confidence levels in the $\Omega_m - h r_d$ contour plane for different tracers corresponding to various $z_{\text{eff}}$ values from the DESI DR2 dataset within the $\Lambda$CDM model. The horizontal lines correspond to the mean values of $\Omega_m$ for each tracer, and the gray band represents the Planck $\Lambda$CDM prediction for $\Omega_m = 0.315 \pm 0.007$.}\label{fig_1}
\end{figure}

Motivated by Fig.~\ref{fig_1}, the LRG1 ($z_{\mathrm{eff}}=0.510$) and LRG3+ELG1
($z_{\mathrm{eff}}=0.934$) BAO points appear in tension with the Planck $\Lambda$CDM value $\Omega_m = 0.315 \pm 0.007$, at about $2.42\sigma$ and $2.60\sigma$, respectively. This trend becomes more prominent when comparing the value of $\Omega_m$ at $z_{\mathrm{eff}} = 0.510$ and $z_{\mathrm{eff}} = 0.934$ from the LRG1 and LRG3+ELG1 datasets, which are in tension with various supernovae compilations, showing discrepancies of $2.06\sigma$, $1.67\sigma$, and $1.80\sigma$ for LRG1 and $2.24\sigma$, $2.51\sigma$, and $2.96\sigma$ for LRG3+ELG1 with Pantheon$^+$, Union3, and DES-SN5YR, respectively. These supernova datasets typically have similarly low effective redshifts around $z_{\mathrm{eff}} \sim 0.3$. See Table~1 and Sec.~4 of \cite{chaudhary2025does} for details.

This behavior is not new, as similar trends have been observed in previous studies. \cite{colgain2024does} (see Table~1 and Figure~5) report the LRG1 discrepancy using the
DESI~DR1 compilation, and note that in DESI~DR1 the LRG2 point also exhibits a significant discrepancy. Similar discrepancies are found in \cite{o2022revealing} using SDSS-IV data
(see Table~3 and Figure~5). More recently, using DESI~DR2, \cite{chaudhary2025does,colgain2025much} show that the LRG1 ($z_{\mathrm{eff}}=0.510$) discrepancy persists, and that a comparable tension is also present for LRG3+ELG1 ($z_{\mathrm{eff}}=0.934$).

Indeed, recent work by \cite{Zheng:2024qzi} has explicitly investigated the role of individual DESI tracers in driving the preference for dynamical dark energy. In particular, they show that removing the LRG1 and LRG2 data points significantly affects the inferred constraints on dark energy parametrizations, indicating that these low-redshift tracers play a disproportionate role in driving deviations from $\Lambda$CDM. This provides further motivation for treating LRG1 and LRG2 on the same footing in our analysis.

\begin{figure}
\centering
\includegraphics[scale=0.55]{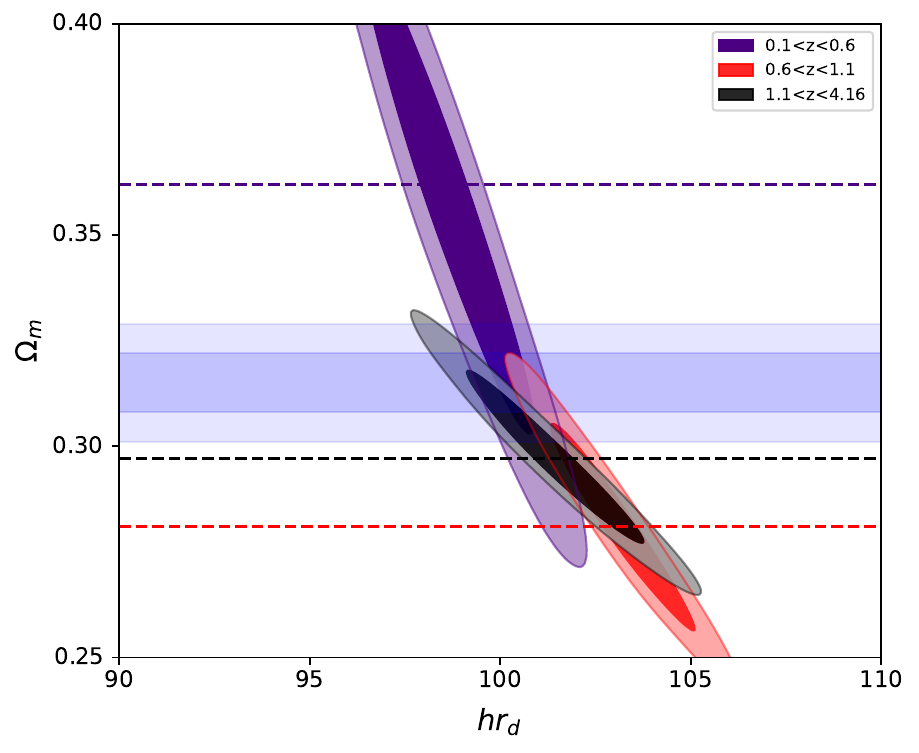}
\caption{The figure shows the posterior distributions in the $\Omega_m - h r_d$ plane using different redshift bins from the DESI DR2 compilation. The blue band represents the Planck $\Lambda$CDM prediction for $\Omega_m$.}\label{fig_2}
\end{figure}

In Fig.~\ref{fig_2}, considering next information from high-redshift bins alone shows a tension of about $1.84\sigma$, indicating that $\Omega_m$ differs between different redshift bins. A similar trend can be observed in the DESI DR1 data, where \cite{colgain2024does} reports a tension of about $2.20\sigma$ when considering next information from high-redshift bins alone. In fact, DESI DR1 shows some improvement; see Table~2 of \cite{chaudhary2025does,colgain2024does} for more details. These findings lead us to investigate the DESI DR2 data beyond the $\Lambda$CDM paradigm. In this study, we extend our analysis beyond the $\Lambda$CDM model and consider several DE models to investigate the effects of DESI BAO tracers beyond $\Lambda$CDM. 

Unlike previous studies, our analysis is based on the DESI DR2 dataset and explores a broader class of dark energy parametrizations, including both phenomenological ($w_0$--$w_a$) models and physically motivated scenarios such as GEDE. We also perform a systematic comparison by removing LRG1 and LRG2 both individually and jointly, allowing us to directly quantify their impact on the inferred dark energy evolution

\begin{figure*}
\begin{subfigure}{.30\textwidth}
\includegraphics[width=\linewidth]{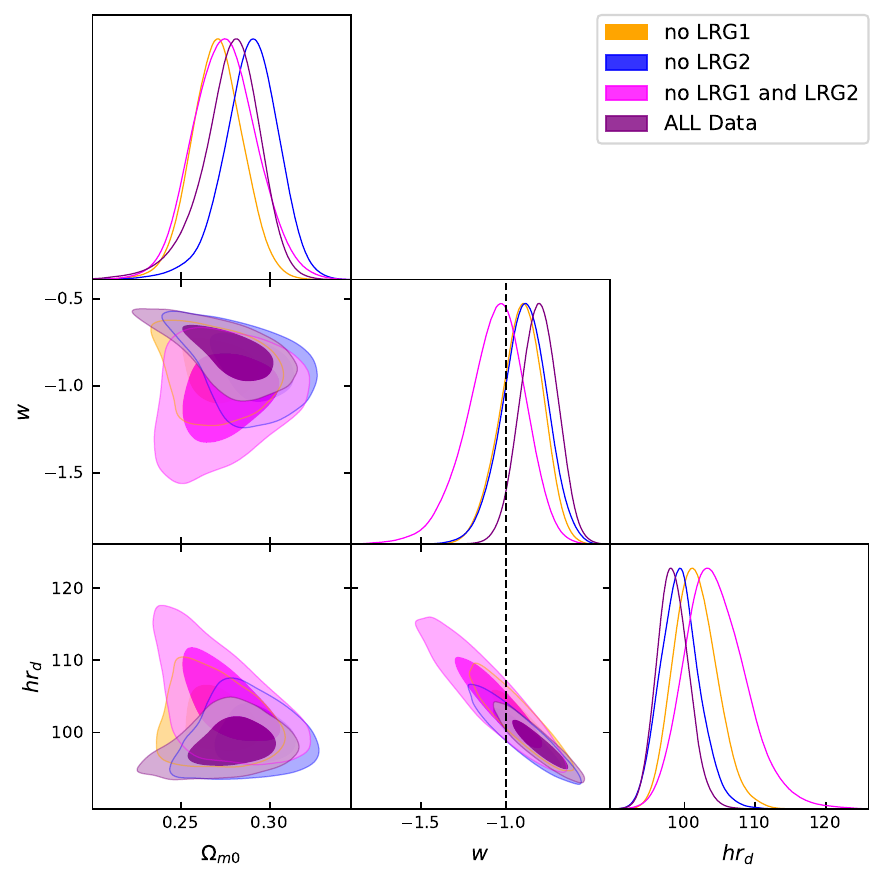}
    \caption{$w$CDM Model}
    \label{fig_3a}
\end{subfigure}
\hfil
\begin{subfigure}{.30\textwidth}
\includegraphics[width=\linewidth]{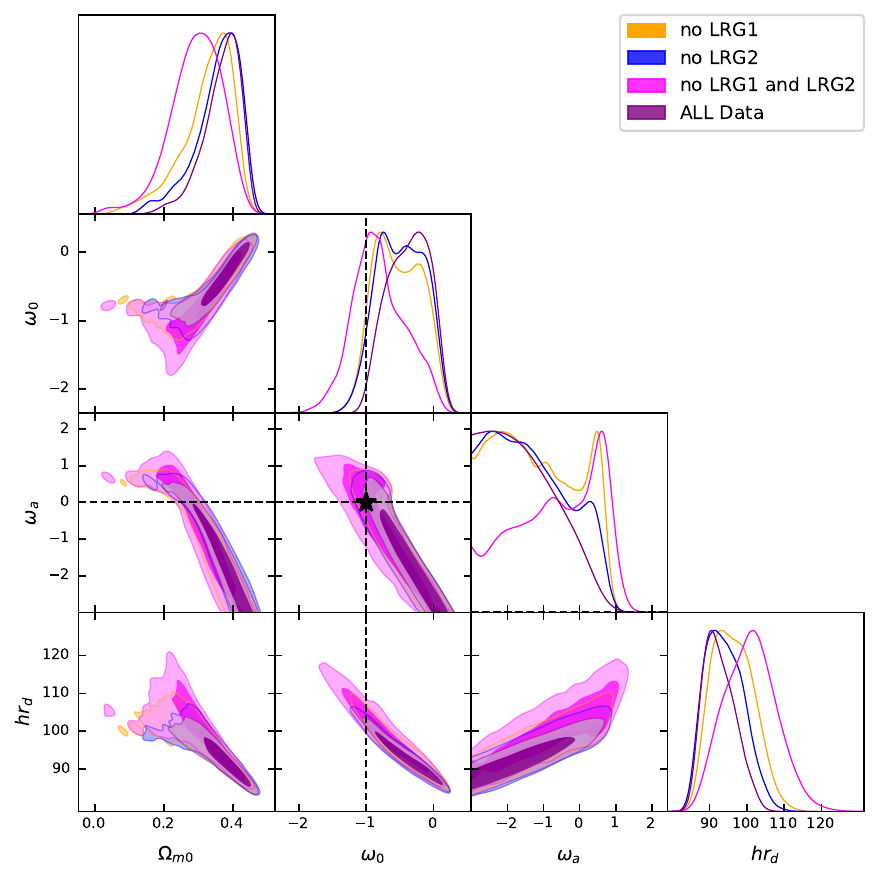}
    \caption{Logarithmic Parameterization}
    \label{fig_3b}    
\end{subfigure}
\hfil
\begin{subfigure}{.30\textwidth}
\includegraphics[width=\linewidth]{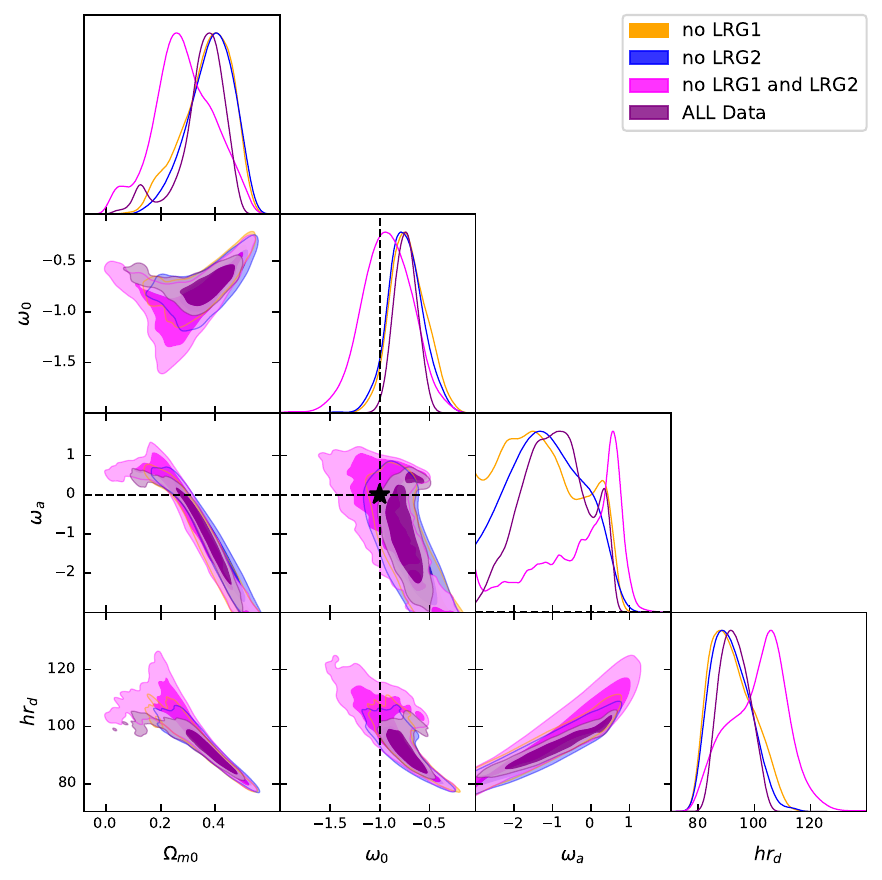}
    \caption{Exponential Parameterization}
    \label{fig_3c}
\end{subfigure}
\hfil
\begin{subfigure}{.30\textwidth}
\includegraphics[width=\linewidth]{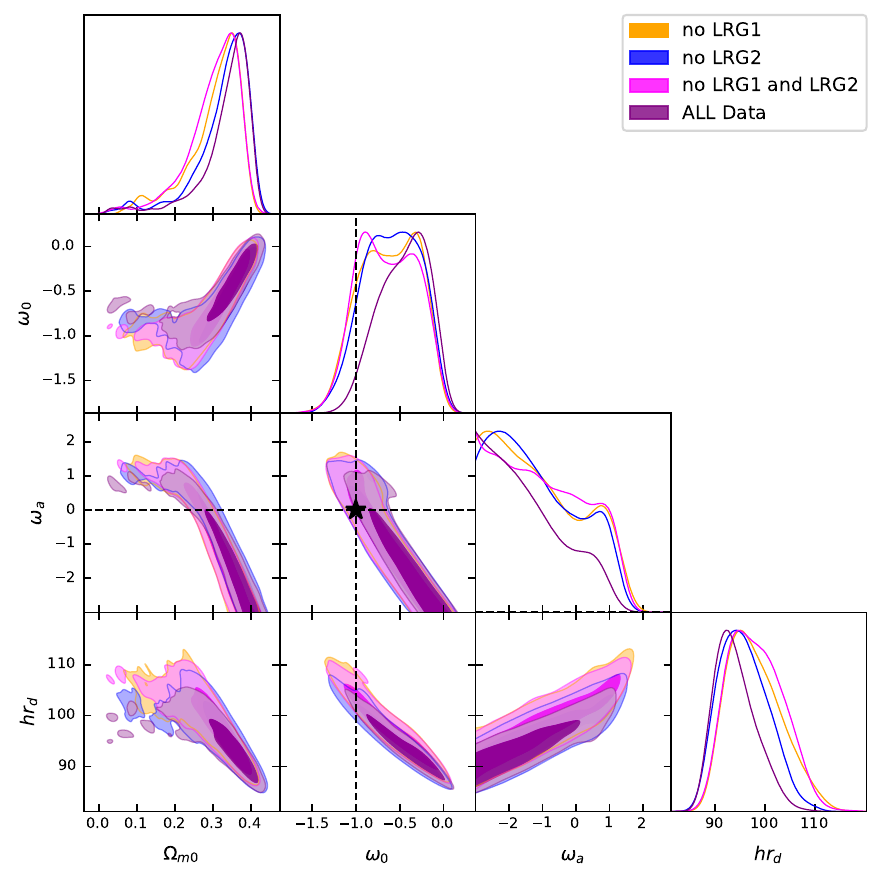}
    \caption{CPL Parameterization}
    \label{fig_3d}
\end{subfigure}
\hfil
\begin{subfigure}{.30\textwidth}
\includegraphics[width=\linewidth]{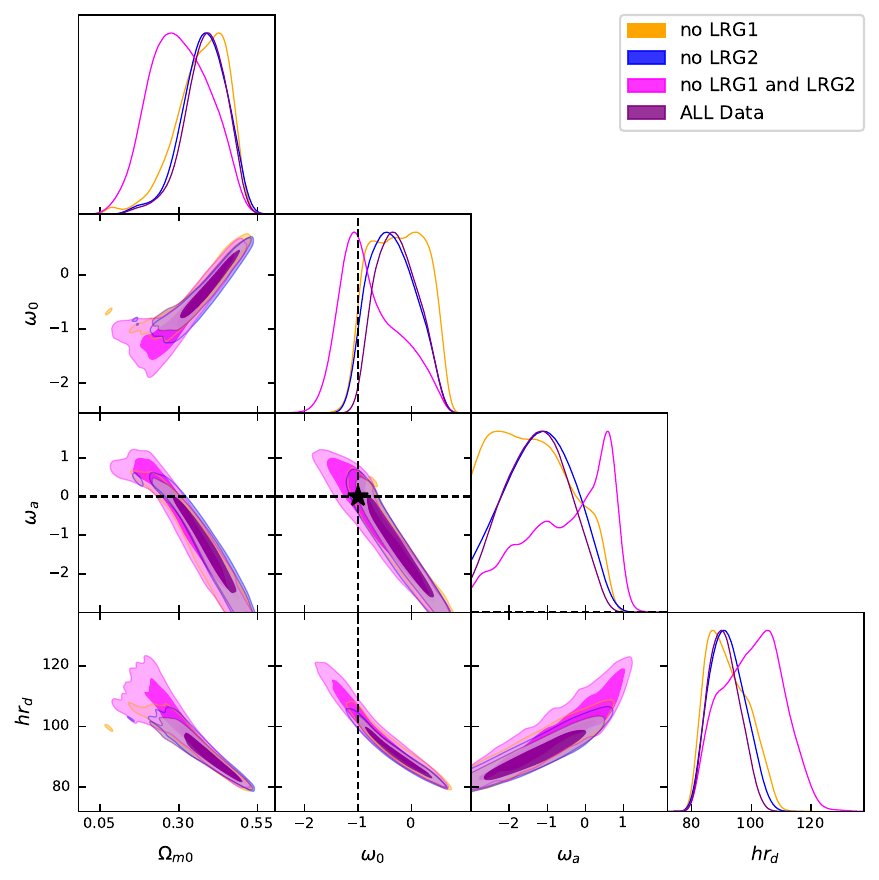}
    \caption{BA Parameterization}
    \label{fig_3e}
\end{subfigure}
\hfil
\begin{subfigure}{.30\textwidth}
\includegraphics[width=\linewidth]{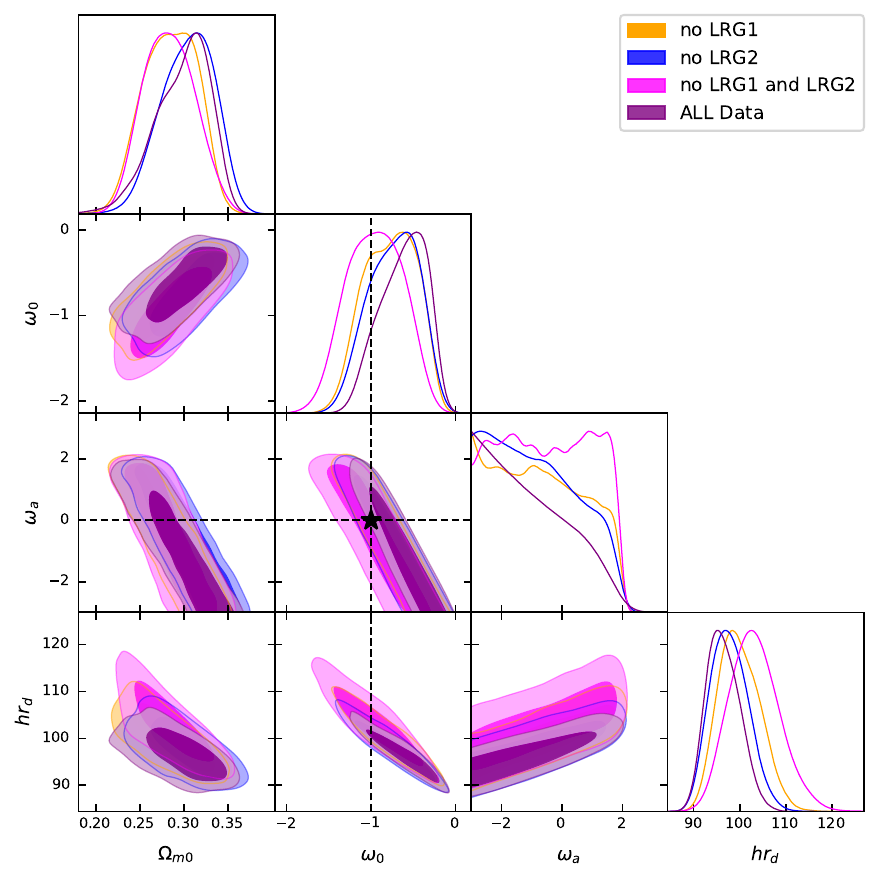}
    \caption{JBP Parameterization}
    \label{fig_3f}
\end{subfigure}
\hfil
\begin{subfigure}{.30\textwidth}
\includegraphics[width=\linewidth]{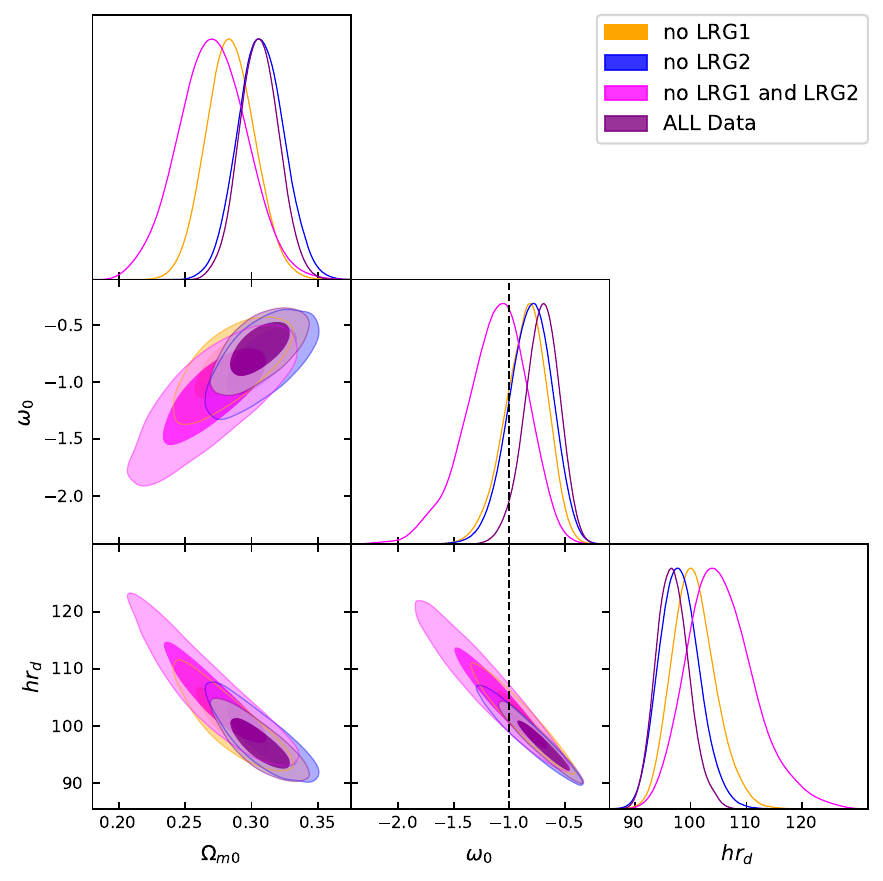}
    \caption{Thawing Parameterization}
    \label{fig_3g}
\end{subfigure}
\hfil
\begin{subfigure}{.30\textwidth}
\includegraphics[width=\linewidth]{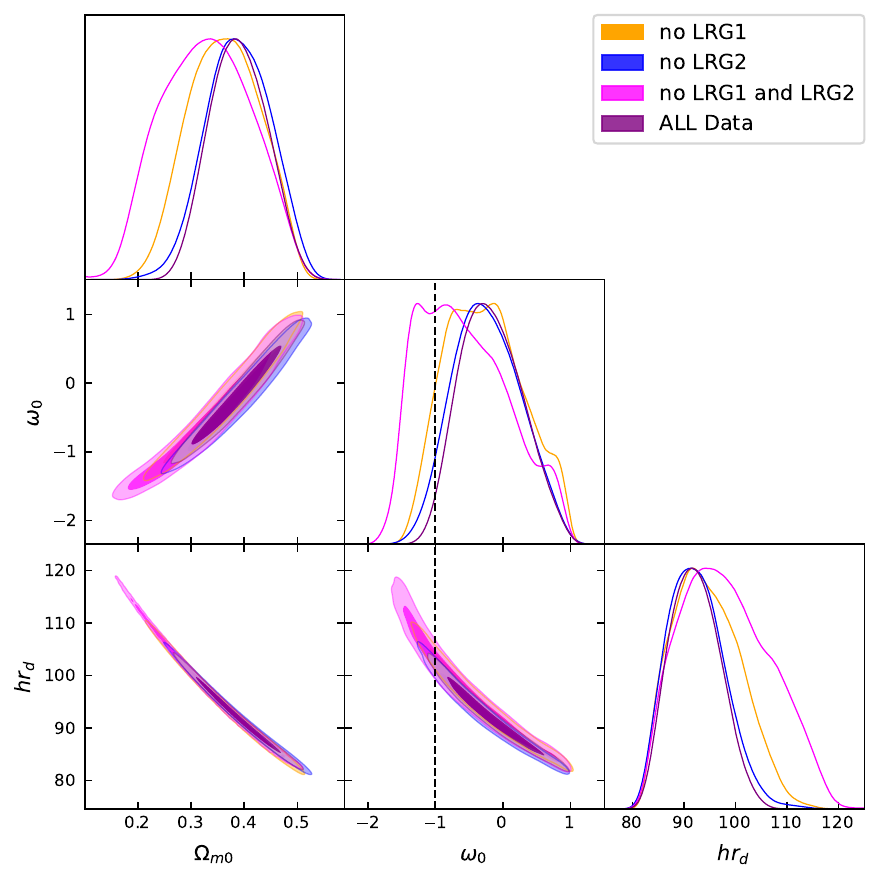}
    \caption{Mirage Parameterization}
    \label{fig_3h}
\end{subfigure}
\hfil
\begin{subfigure}{.30\textwidth}
\includegraphics[width=\linewidth]{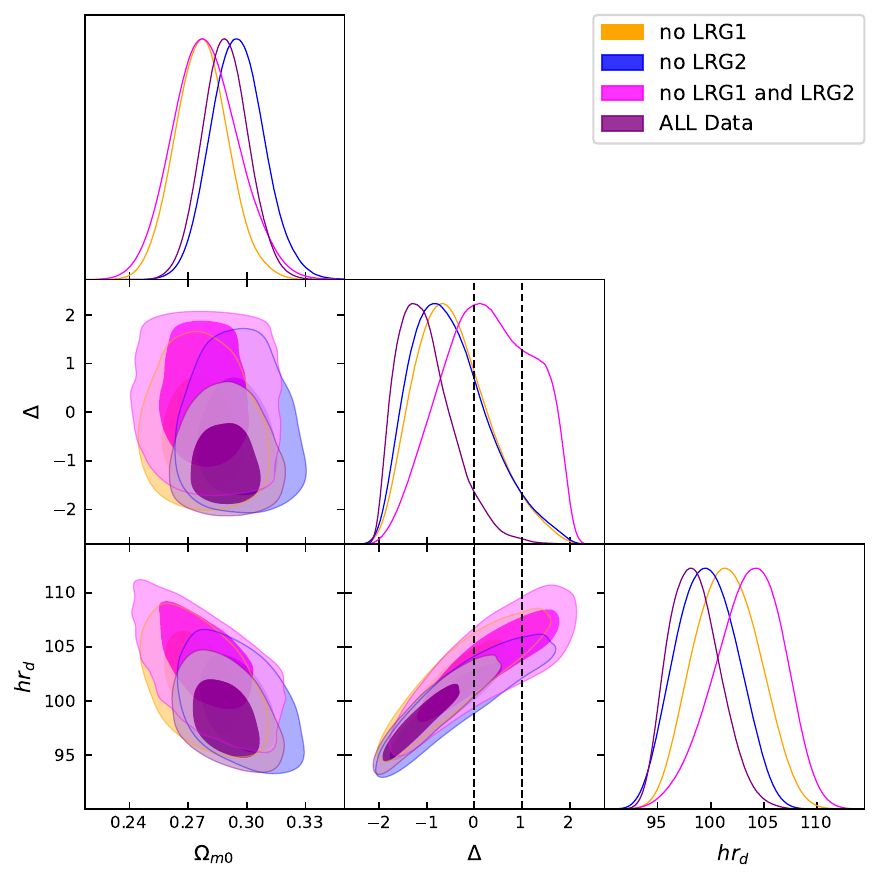}
    \caption{GEDE Parameterization}
    \label{fig_3i}
\end{subfigure}
\caption{The figure shows the posterior distributions $w$CDM, Logarithmic, Exponential, CPL, BA, JBP, Thawing, Mirage, and GEDE models at 68\% (1$\sigma$) and 95\% (2$\sigma$) confidence, using no LRG1, no LRG2, no LRG1 \& LRG2, and the full DESI DR2 sample.}\label{fig_3}
\end{figure*}

In Fig.~\ref{fig_3} shows the triangle plot of the different DE models using the DESI DR2 measurements (excluding the BGS datapoint), for no LRG1, no LRG2, no LRG1 \& LRG2, and the full DESI DR2 sample. The diagonal panels show the 1D marginalized posterior distributions for each parameter. The off-diagonal panels show the 2D marginalized confidence contours at 68\% and 95\% intervals. Table~\ref{tab_3} shows the numerical values obtained for each model using MCMC analysis.

Fig.~\ref{fig_3a} shows the constraints on the $w$CDM model. When the LRG1 and LRG2 datasets are included, the inferred $w$ deviates from the $\Lambda$CDM prediction ($w=-1$). Conversely, removing LRG1 and LRG2 fully restores the $\Lambda$CDM concordance. 

Figs.~\ref{fig_3b}, \ref{fig_3c}, \ref{fig_3d}, \ref{fig_3e}, and \ref{fig_3f} show the constraints on the Logarithmic, Exponential, CPL, BA, and JBP redshift parameterizations. In each case, it can be observed that when the LRG1 and LRG2 data points are included alternately in the analysis, the predicted value $w_0>-1$ shows a deviation from the $(w_0 =-1,w_a=0)$ corresponds to the $\Lambda$CDM model in $w_0-w_a$ plane, and it can be observed that $\omega_a$ also shifts to large negative values, extending beyond the previous limits to accommodate $w_0>-1$. This suggests that the evolving DE in the DESI DR2 data could be driven by the LRG1 and LRG2 datapoints, particularly at low redshift ($z \lesssim 0.3$). This behavior is consistent with the findings of \cite{Zheng:2024qzi}, where the removal of LRG1 and LRG2 similarly reduces the apparent preference for dynamical dark energy, reinforcing the interpretation that these tracers play a key role in driving the signal.

Figs.~\ref{fig_3g} and \ref{fig_3h} show the constraints on the Calibrated Thawing and Calibrated Mirage models using DESI DR2 measurements. Similar effects of the LRG1 and LRG2 datapoints are observed for $\omega_0$: when LRG1 and LRG2 are included alternately, the inferred value satisfies $w_0>-1$ ,which in turn drives $w_a<0$. For the Mirage model, we again find  $w_0>-1$ and, consequently, $w_a<0$.

Fig.~\ref{fig_3i} shows the constraints on the GEDE model. Similar effects of the LRG1 and LRG2 datapoints are also observed here: when the LRG1 and LRG2 datapoints are included alternately, the model predicts $\Delta = -1.010$, which lies far from the $\Lambda$CDM point ($\Delta = 0$). In contrast, without the LRG1 and LRG2 datapoints, GEDE predicts $\Delta = -0.300$, much closer to $\Lambda$CDM. In both cases, the negative value of $\Delta$ indicates an injection of energy at earlier redshifts. These results highlight that the LRG1 and LRG2 tracers play a crucial role in driving the preference for dynamical dark energy in the DESI DR2 BAO dataset.

In addition to the effects of the LRG1-2 tracers, an interesting feature is also evident: a negative correlation between $\Omega_m$ and $h r_d$. This anti-correlation has been observed previously in several independent analyses. Specifically, \cite{colgain2024putting,dainotti2022evolution} have reported this trend using Observational Hubble Data (OHD); \cite{o2022revealing,dainotti2021hubble,colgain2024putting,jia2023evidence,pasten2023testing,malekjani2024redshift,wagner2022solving,hu2022revealing,dainotti2023hubble} using Type Ia supernovae (SNe Ia); \cite{colgain2024putting,krishnan2020there} by combining OHD and SNe Ia; \cite{dainotti2022gamma,Bargiacchi:2024srw} using Gamma-Ray Bursts (GRBs); \cite{o2022revealing,colgain2024putting,risaliti2019cosmological,lusso2020quasars,dainotti2023quasars,dainotti2022quasar,bargiacchi2023gamma,pourojaghi2022can} using standardizable QSOs; and similar patterns have been discussed in studies of strong-lensing time delays in lensed QSOs \cite{wong2020h0licow,shajib2020strides,millon2020tdcosmo} and lensed SNe Ia \cite{kelly2023constraints,pascale2025sn}.

It is worth noting that each tracer provides only a limited set of observables, such as $D_H/r_d$ and $D_M/r_d$, while the fitted models involve a larger number of parameters. This can lead to an underconstrained setup, increasing the risk of overfitting and posterior instability, particularly when individual redshift bins are analyzed. As a result, the apparent anomaly in $\Omega_m$ may reflect statistical artifacts rather than genuine physical constraints. Motivated by this, we go beyond the DESI BAO-only analysis in the following sections by incorporating different Type~Ia supernova samples and a CMB compressed likelihood.

\begin{table*}
\setlength{\tabcolsep}{12pt}
\begin{tabular}{lcccccccccc}
\hline
\textbf{Dataset/Models} & $\Omega_{m0}$ & $w$ or $w_0$ & $w_a$ & $\Delta$ & $h\,r_d$ \\
\hline
\textbf{$\Lambda$CDM} \\
No LRG1 & $0.275{\pm0.013}$ & --- & --- & --- & $101.4{\pm1.2}$ \\
No LRG2 & $0.296{\pm0.013}$ & --- & --- & --- & $103.4{\pm1.3}$ \\
No LRG1 \& LRG2 & $0.281_{-0.019}^{+0.016}$ & --- & --- & --- & $102.8{\pm1.9}$ \\
DESI DR2 & $0.289{\pm0.011}$ & --- & --- & --- & $102.1{\pm1.0}$ \\
\hline
\textbf{$w$CDM} \\
No LRG1 & $0.270{\pm0.015}$ & $-0.910_{-0.110}^{+0.140}$ & --- & --- & $101.6_{-3.3}^{+2.7}$ \\
No LRG2 & $0.289_{-0.014}^{+0.017}$ & $-0.890_{-0.120}^{+0.140}$ & --- & --- & $99.4_{-3.0}^{+2.4}$ \\
No LRG1 \& LRG2 & $0.275{\pm0.018}$ & $-1.07_{-0.150}^{+0.190}$ & --- & --- & $104.6_{-4.9}^{+3.6}$ \\
DESI DR2 & $0.277_{-0.013}^{+0.019}$ & $-0.810{\pm0.110}$ & --- & --- & $98.4_{-2.4}^{+2.1}$ \\
\hline
\textbf{Logarithmic} \\
No LRG1 & $0.323^{+0.092}_{-0.038}$ & $-0.530{\pm0.350}$ & $-1.20{\pm1.10}$ & --- & $96.3_{-6.5}^{+4.9}$ \\
No LRG2 & $0.355^{+0.081}_{-0.035}$ & $-0.480{\pm0.340}$ & $-1.32_{-1.60}^{+0.62}$ & --- & $94.1_{-6.1}^{+4.4}$ \\
No LRG1 \& LRG2 & $0.293^{+0.085}_{-0.060}$ & $-0.810_{-0.460}^{+0.340}$ & $-0.630_{-0.78}^{+1.60}$ & --- & $101.0{\pm7.0}$ \\
DESI DR2 & $0.366^{+0.066}_{-0.033}$ & $-0.370_{-0.280}^{+0.380}$ & $-1.530_{-1.400}^{+0.053}$ & --- & $92.7_{-5.2}^{+3.4}$ \\
\hline
\textbf{Exponential} \\
No LRG1 & $0.370^{+0.120}_{-0.061}$ & $-0.710_{-0.200}^{+0.160}$ & $-1.30_{-1.50}^{+1.70}$ & --- & $92.1_{-10.0}^{+5.5}$ \\
No LRG2 & $0.383^{+0.100}_{-0.066}$ & $-0.740{\pm0.180}$ & $-1.16{\pm0.94}$ & --- & $91.6_{-8.3}^{+5.7}$ \\
No LRG1 \& LRG2 & $0.280^{+0.120}_{-0.100}$ & $-0.930{\pm0.260}$ & $-0.370_{-0.57}^{+1.30}$ & --- & $102_{-9.4}^{+11.0}$ \\
DESI DR2 & $0.349^{+0.040}_{-0.095}$ & $-0.740{\pm0.110}$ & $-0.950_{-0.900}^{+1.300}$ & --- & $93.0_{-6.0}^{+4.8}$ \\
\hline
\textbf{CPL} \\
No LRG1 & $0.303_{-0.027}^{+0.083}$ & $-0.620_{-0.300}^{+0.410}$ & $-1.00{\pm1.30}$ & --- & $97.8_{-6.7}^{+4.1}$ \\
No LRG2 & $0.325_{-0.024}^{+0.080}$ & $-0.590_{-0.310}^{+0.370}$ & $-1.10_{-1.80}^{+2.00}$ & --- & $95.8_{-5.9}^{+4.0}$ \\
No LRG1 \& LRG2 & $0.275{\pm0.065}$ & $-0.950_{-0.470}^{+0.340}$ & $-0.270_{-0.93}^{+2.00}$ & --- & $103.4{\pm7.1}$ \\
DESI DR2 & $0.333_{-0.019}^{+0.072}$ & $-0.460_{-0.240}^{+0.350}$ & $-1.410_{-1.600}^{+0.500}$ & --- & $94.1_{-5.1}^{+3.0}$ \\
\hline
\textbf{BA} \\
No LRG1 & $0.373_{-0.055}^{+0.099}$ & $-0.240{\pm0.490}$ & $-1.37_{-1.40}^{+0.66}$ & --- & $92.2_{-9.0}^{+5.3}$ \\
No LRG2 & $0.380_{-0.061}^{+0.078}$ & $-0.350_{-0.520}^{+0.380}$ & $-1.18{\pm0.88}$ & --- & $92.1_{-6.9}^{+5.2}$ \\
No LRG1 \& LRG2 & $0.303_{-0.110}^{+0.092}$ & $-0.750_{-0.710}^{+0.400}$ & $-0.520_{-0.760}^{+1.400}$ & --- & $100.6{\pm9.7}$ \\
DESI DR2 & $0.383_{-0.054}^{+0.075}$ & $-0.270_{-0.470}^{+0.350}$ & $-1.260{\pm0.810}$ & --- & $90.9_{-6.1}^{+4.5}$ \\
\hline
\textbf{JBP} \\
No LRG1 & $0.284_{-0.029}^{+0.034}$ & $-0.770_{-0.300}^{+0.350}$ & $-0.700_{-2.200}^{+1.000}$ & --- & $99.9_{-5.0}^{+4.0}$ \\
No LRG2 & $0.304_{-0.027}^{+0.034}$ & $-0.730_{-0.270}^{+0.350}$ & $-0.840_{-2.100}^{+0.970}$ & --- & $97.8_{-4.5}^{+3.7}$ \\
No LRG1 \& LRG2 & $0.283_{-0.032}^{+0.028}$ & $-0.950{\pm0.330}$ & $-0.500{\pm1.400}$ & --- & $103.2_{-5.9}^{+4.8}$ \\
DESI DR2 & $0.297_{-0.024}^{+0.038}$ & $-0.620_{-0.210}^{+0.340}$ & $-1.050_{-1.900}^{+0.710}$ & --- & $96.5_{-4.1}^{+3.2}$ \\
\hline
\textbf{Thawing} \\
No LRG1 & $0.284{\pm0.018}$ & $-0.850_{-0.170}^{+0.210}$ & --- & --- & $100.7_{-4.3}^{+3.4}$ \\
No LRG2 & $0.307{\pm0.017}$ & $-0.810_{-0.180}^{+0.210}$ & --- & --- & $98.1_{-3.9}^{+3.2}$ \\
No LRG1 \& LRG2 & $0.271{\pm0.026}$ & $-1.130_{-0.240}^{+0.310}$ & --- & --- & $105.7_{-6.8}^{+4.9}$ \\
DESI DR2 & $0.306{\pm0.015}$ & $-0.710_{-0.140}^{+0.170}$ & --- & --- & $96.8_{-3.2}^{+2.6}$ \\
\hline
\textbf{Mirage} \\
No LRG1 & $0.363{\pm0.066}$ & $-0.260_{-0.710}^{+0.480}$ & --- & --- & $94.4_{-7.8}^{+5.7}$ \\
No LRG2 & $0.389{\pm0.060}$ & $-0.230_{-0.560}^{+0.470}$ & --- & --- & $92.4_{-6.3}^{+4.6}$ \\
No LRG1 \& LRG2 & $0.328{\pm0.085}$ & $-0.570_{-0.880}^{+0.450}$ & --- & --- & $98.5_{-11.0}^{+6.8}$ \\
DESI DR2 & $0.389{\pm0.053}$ & $-0.170_{-0.530}^{+0.420}$ & --- & --- & $92.2_{-5.4}^{+4.5}$ \\
\hline
\textbf{GEDE} \\
No LRG1 & $0.277{\pm0.014}$ & --- & --- & $-0.450_{-0.930}^{+0.600}$ & $101.5{\pm3.0}$ \\
No LRG2 & $0.295{\pm0.014}$ & --- & --- & $-0.500_{-1.000}^{+0.570}$ & $99.6{\pm2.8}$ \\
No LRG1 \& LRG2 & $0.280_{-0.019}^{+0.015}$ & --- & --- & $-0.300_{-0.890}^{+1.100}$ & $103.5_{-2.8}^{+3.7}$ \\
DESI DR2 & $0.289{\pm0.012}$ & --- & --- & $-1.010_{-0.750}^{+0.360}$ & $98.5_{-2.0}^{+2.0}$ \\
\hline
\end{tabular}
\caption{This table presents the numerical values obtained using DESI DR2 measurements for the $\Lambda$CDM, $w$CDM, Logarithmic, Exponential, CPL, BA, JBP, Thawing, Mirage, and GEDE models, with and without the inclusion of the LRG1 datapoint at the 68\% (1$\sigma$) confidence level.}
\label{tab_3}
\end{table*}

\begin{table*}
\setlength{\tabcolsep}{5pt}
\resizebox{\textwidth}{!}{%
\begin{tabular}{lcccccccccc}
\hline
\textbf{Dataset/Models} & $H_0$ & $\Omega_{m0}$ & $w$ or $w_0$ & $w_a$ & $\Delta$ & $\ln\mathcal{Z}$ & $\ln \mathcal{B}_{\mathrm{\text{DE Model}},\Lambda\mathrm{CDM}}$ & $\Delta{\chi^{2}_{\text{MAP}}}$ & $N\sigma$ \\
\hline

\textbf{$\Lambda$CDM} \\
CMB + DESI DR2 & $68.19{\pm0.46}$ & $0.3035{\pm0.0056}$ & --- & --- & --- & -20.01 & 0 & 0 & 0 \\
CMB + DESI DR2 + Pantheon$^+$ & $67.95{\pm0.48}$ & $0.3064{\pm0.0058}$ & --- & --- & --- & -723.11 & 0 & 0 & 0 \\
CMB + DESI DR2 + DES-Dovekie & $67.90{\pm0.43}$ & $0.3072{\pm0.0052}$  & --- & --- & --- & -837.86 & 0 & 0 & 0 \\
CMB + DESI DR2 + Union3 & $67.97_{-0.50}^{+0.44}$ & $0.3064{\pm0.0057}$  & --- & --- & --- & -33.84 & 0 & 0 & 0 \\
\hline

\textbf{$w$CDM} \\
CMB + DESI DR2 & $68.77{\pm0.99}$ & $0.2990{\pm0.0085}$ & $-1.037{\pm0.041}$ & --- & --- & -23.33 & -3.32 & 0.29 & 0  \\
CMB + DESI DR2 + Pantheon$^+$ & $67.57{\pm0.65}$ & $0.3085{\pm0.0065}$ & $-0.986{\pm0.026}$ & --- & --- & -726.58 & -3.47 & 0.22 & 0  \\
CMB + DESI DR2 + DES-Dovekie & $67.46{\pm0.57}$ & $0.3094{\pm0.0058}$  & $-0.983{\pm0.024}$ & --- & --- & -841.51 & -3.65 & -0.02 & 0.14 \\
CMB + DESI DR2 + Union3 & $67.37{\pm0.76}$ & $0.3103{\pm0.0072}$  & $-0.979{\pm0.031}$ & --- & --- & -37.40 & -3.56 & 0.11 & 0 \\
\hline

\textbf{CPL} \\
CMB + DESI DR2 & $62.6_{-1.9}^{+1.4}$ & $0.366_{-0.018}^{+0.023}$  & $-0.33_{-0.14}^{+0.23}$ & $-2.07_{-0.80}^{+0.34}$ & --- & -18.62 & 1.39 & -4.62 & 1.65 \\
CMB + DESI DR2 + Pantheon$^+$ & $67.30{\pm0.63}$ & $0.3137{\pm0.0067}$  & $-0.858{\pm0.058}$ & $-0.58{\pm 0.24}$ & --- & -725.19 & -2.08 & -2.71 & 1.13 \\
CMB + DESI DR2 + DES-Dovekie & $67.16{\pm0.59}$ & $0.3156{\pm0.0065}$  & $-0.821{\pm0.059}$ & $-0.73_{-0.24}^{+0.27}$ & --- & -838.75 & -0.89 & -4.21 & 1.55 \\
CMB + DESI DR2 + Union3 & $65.56{\pm0.86}$ & $0.3320{\pm0.0096}$  & $-0.662{\pm0.091}$ & $-1.15{\pm 0.33}$ & --- & -32.03 & 1.81 & -6.46 & 2.06 \\
\hline

\textbf{Logarithmic} \\
CMB + DESI DR2 & $62.9{\pm2.1}$ & $0.365_{-0.028}^{+0.024}$  & $-0.42_{-0.25}^{+0.21}$ & $-1.39_{-0.47}^{+0.61}$ & --- & -19.42 & 0.59 & -4.40 & 1.59 \\
CMB + DESI DR2 + Pantheon$^+$ & $67.38{\pm0.68}$ & $0.3138{\pm0.0072}$  & $-0.871{\pm0.051}$ & $-0.39_{-0.16}^{+0.17}$ & --- & -724.77 & -1.66 & -3.41 & 1.34 \\
CMB + DESI DR2 + DES-Dovekie & $67.28{\pm0.59}$ & $0.3151{\pm0.0064}$  & $-0.849{\pm0.050}$ & $-0.46{\pm 0.17}$ & --- & -838.76 & -0.90 & -4.63 & 1.65 \\
CMB + DESI DR2 + Union3 & $67.76{\pm0.83}$ & $0.3308_{-0.0099}^{+0.0084}$  & $-0.710{\pm0.078}$ & $-0.75_{-0.21}^{+0.24}$ & --- & -32.25 & 1.59 & -6.76 & 2.12 \\
\hline

\textbf{Exponential} \\
CMB + DESI DR2 & $61.7{\pm 2.3}$ & $0.379_{-0.031}^{+0.028}$  & $-0.72{\pm 0.11}$ & $-1.06_{-0.36}^{+0.41}$ & --- & -19.84 & 0.17 & -4.57 & 1.64 \\
CMB + DESI DR2 + Pantheon$^+$ & $67.36{\pm0.64}$ & $0.3133{\pm0.0068}$  & $-0.971{\pm0.027}$ & $-0.230_{-0.098}^{+0.11}$ & --- & -726.23 & -3.12 & -2.91 & 1.19 \\
CMB + DESI DR2 + DES-Dovekie & $67.22{\pm 0.59}$ & $0.3151{\pm0.0065}$  & $-0.965{\pm0.025}$ & $-0.29{\pm 0.11}$ & --- & -839.22 & -1.36 & -4.38 & 1.59 \\
CMB + DESI DR2 + Union3 & $65.63{\pm0.87}$ & $0.3314{\pm0.0097}$  & $-0.898{\pm0.038}$ & $-0.48{\pm0.14}$ & --- & -32.77 & 1.07 & -6.56 & 2.08 \\
\hline

\textbf{JBP} \\
CMB + DESI DR2 & $65.8_{-2.0}^{+1.1}$ & $0.328_{-0.012}^{+0.020}$  & $-0.627_{-0.086}^{+0.23}$ & $<-1.68$ & --- & -21.88 & -1.87 & -2.62 & 1.10 \\
CMB + DESI DR2 + Pantheon$^+$ & $67.38{\pm0.65}$ & $0.3120{\pm0.0066}$  & $-0.847{\pm0.080}$ & $-0.85{\pm0.49}$ & --- & -725.83 & -2.72 & -1.79 & 0.83 \\
CMB + DESI DR2 + DES-Dovekie & $67.10{\pm0.61}$ & $0.3151{\pm0.0065}$  & $-0.774{\pm0.085}$ & $-1.29{\pm0.52}$ & --- & -839.12 & -1.26 & -3.57 & 1.38 \\
CMB + DESI DR2 + Union3 & $65.65{\pm0.88}$ & $0.3292{\pm0.0093}$  & $-0.599_{-0.088}^{+0.12}$ & $-2.06_{-0.73}^{+0.39}$ & --- & -33.09 & 0.75 & -5.36 & 1.82 \\
\hline

\textbf{BA} \\
CMB + DESI DR2 & $62.4{\pm 2.1}$ & $0.370_{-0.028}^{+0.023}$  & $-0.39_{-0.23}^{+0.20}$ & $-1.01_{-0.32}^{+0.39}$ & --- & -19.43 & 0.58 & -4.65 & 1.66 \\
CMB + DESI DR2 + Pantheon$^+$ & $67.37{\pm0.64}$ & $0.3134{\pm0.0067}$  & $-0.881{\pm0.048}$ & $-0.26{\pm 0.11}$ & --- & -725.39 & -2.28 & -3.17 & 1.27 \\
CMB + DESI DR2 + DES-Dovekie & $67.25{\pm0.59}$ & $0.3149{\pm0.0063}$ & $-0.856{\pm0.048}$ & $-0.31{\pm 0.12}$ & --- & -838.94 & -1.08 & -4.37 & 1.59 \\
CMB + DESI DR2 + Union3 & $65.74{\pm0.88}$ & $0.3306{\pm0.0097}$ & $-0.722{\pm0.078}$ & $-0.51{\pm 0.15}$ & --- & -32.44 & -1.40 & -6.61 & 2.09 \\
\hline

\textbf{Thawing} \\
CMB + DESI DR2 & $68.5{\pm1.5}$ & $0.304{\pm0.013}$ & $-0.956{\pm0.090}$ & --- & --- & -20.01 & 0.00 & -2.71 & 1.65 \\
CMB + DESI DR2 + Pantheon$^+$ & $68.06{\pm0.65}$ & $0.3077{\pm0.0064}$ & $-0.930{\pm0.040}$ & --- & --- & -722.24 & 0.87 & -3.83 & 1.96 \\
CMB + DESI DR2 + DES-Dovekie & $67.94{\pm0.62}$ & $0.3087{\pm0.0064}$  & $-0.921{\pm0.037}$ & --- & --- & -836.68 & 1.18 & -4.42 & 2.10 \\
CMB + DESI DR2 + Union3 & $67.23{\pm0.83}$ & $0.3151{\pm0.0081}$  & $-0.887{\pm0.052}$ & --- & --- & -32.30 & 1.54 & -4.94 & 2.22 \\
\hline

\textbf{Mirage} \\
CMB + DESI DR2 & $68.34_{-0.78}^{+0.94}$ & $0.3079_{-0.011}^{+0.0091}$ & $-0.87_{-0.13}^{+0.10}$ & --- & --- & -19.25 & 0.76 & -2.98 & 1.73\\
CMB + DESI DR2 + Pantheon$^+$ & $68.43{\pm0.54}$ & $0.3071{\pm0.0065}$ & $-0.889{\pm0.058}$ & --- & --- &  -721.56 & 1.55 & -3.93 & 1.98 \\
CMB + DESI DR2 + DES-Dovekie & $68.22{\pm0.51}$ & $0.3095{\pm0.0061}$  & $-0.856{\pm0.059}$ & --- & --- & -835.67 & 2.19 & -5.00 & 2.24 \\
CMB + DESI DR2 + Union3 & $67.61{\pm0.74}$ & $0.3166{\pm0.0088}$  & $-0.767{\pm0.095}$ & --- & --- & -30.89 & 2.95 & -5.37 & 2.32 \\
\hline

\textbf{GEDE} \\
CMB + DESI DR2 & $68.8{\pm1.0}$ & $0.2994{\pm0.0087}$ & --- & --- & $0.13_{-0.30}^{+0.26}$ & -21.78 & -1.77 & -0.64 & 0.80 \\
CMB + DESI DR2 + Pantheon$^+$ & $67.69{\pm0.63}$ & $0.3084{\pm0.0061}$  & --- & --- & $-0.17{\pm0.17}$ & -724.38 & -1.27 & -0.95 & 0.97 \\
CMB + DESI DR2 + DES-Dovekie & $67.58{\pm0.58}$ & $0.3093{\pm0.0059}$  & --- & --- & $-0.20{\pm0.16}$ & -839.33 & -1.47 & -1.24 & 1.11 \\
CMB + DESI DR2 + Union3 & $67.43{\pm0.81}$ & $0.3105{\pm0.0077}$  & --- & --- & $-0.24{\pm0.22}$ & -35.39 & -1.55 & -1.15 & 1.07 \\
\hline
\end{tabular}
}
\caption{This table presents the numerical values obtained for the $\Lambda$CDM, $w$CDM, Logarithmic, Exponential, CPL, BA, JBP, Thawing, Mirage, and GEDE models at the 68\% (1$\sigma$) confidence level, using DESI DR2 measurements combined with different SNe Ia compilations (Pantheon$^+$, DES-Dovekie, Union3) and CMB compressed likelihood}
\label{tab_4}
\end{table*}

\subsection{Evidence for evolving dark energy using DESI DR2 BAO with other datasets}
In the previous section, we constrained each DE model using only the DESI DR2 dataset. We found that, across all models, the posterior distributions in the $w_0$–$w_a$ plane indicate that $w_a$ remains poorly constrained when relying solely on DESI DR2 data. To improve these constraints and break parameter degeneracies, we now incorporate additional datasets, combining DESI DR2 with various Type~Ia supernova compilations and CMB measurements.

Table~\ref{tab_4} presents the numerical values obtained for each cosmological model using MCMC analysis. Here, we compared the predicted values of the Hubble parameter $h$ and the matter density $\Omega_m$ from each model with the $\Lambda$CDM model using CMB + DESI DR2 data, both alone and in combination with Type Ia supernova datasets (Pantheon$^+$, DES-Dovekie, Union3). Using $\Lambda$CDM as the reference model, we find that most DE parameterizations show weak evidence in both $h$ and $\Omega_m$. In particular, the $w$CDM, Thawing, Mirage, and GEDE models show deviations that remain below the $1\sigma$ level across all dataset combinations, indicating consistency with the $\Lambda$CDM predictions. For the CPL, BA, and JBP parameterizations, the deviations in $h$ and $\Omega_m$ increase but remain below the $2\sigma$ level for all datasets. The largest deviations are observed in the Logarithmic and Exponential models, where tensions in both parameters approach or exceed the $2\sigma$ level for certain dataset combinations.

\begin{figure}
\centering
\includegraphics[scale=0.4]{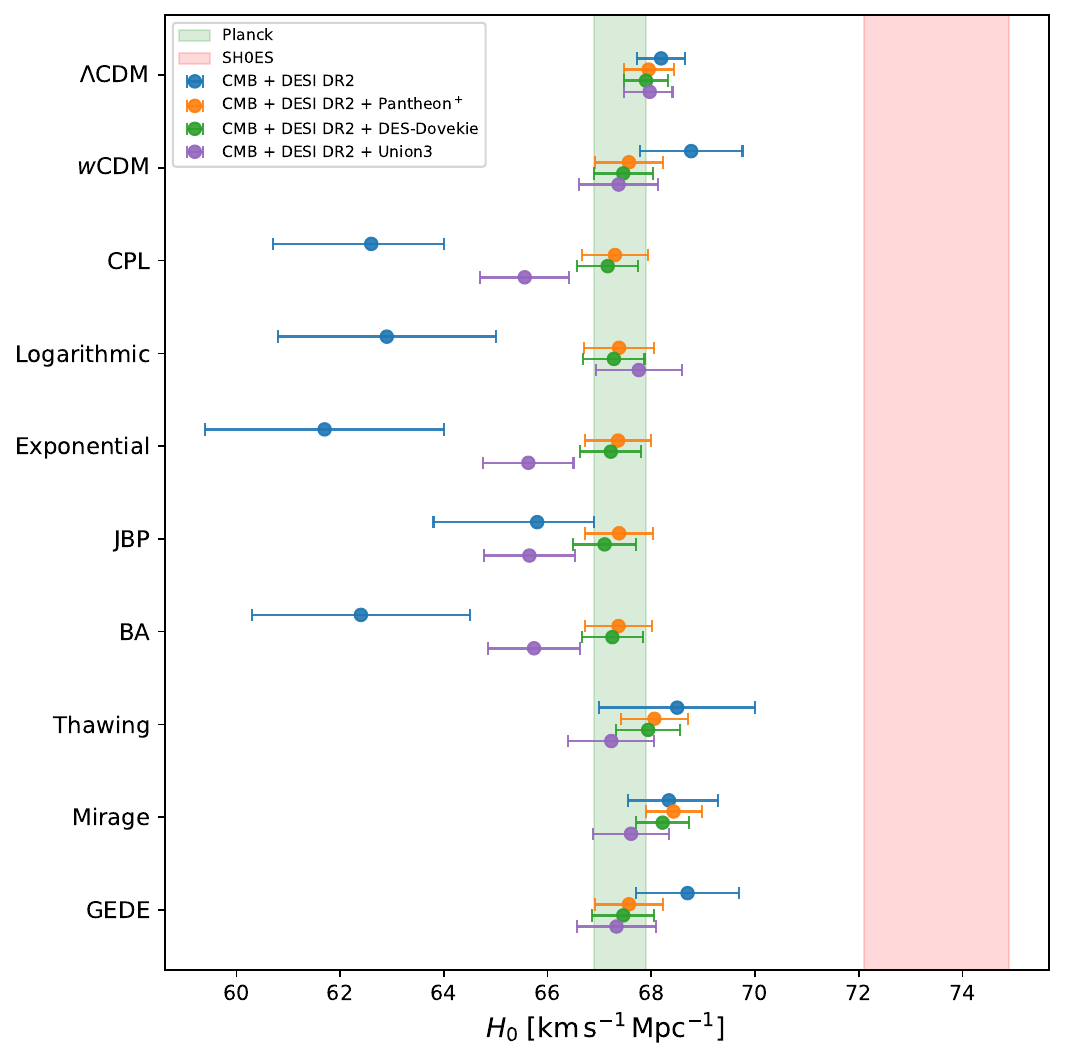}
\caption{The figure shows the comparison of the inferred value of the $h$ for $\Lambda$CDM, $w$CDM, Logarithmic, Exponential, CPL, BA, JBP, Thawing, Mirage, and GEDE models. The green vertical band represents the Planck value $H_0 = 67.4 \pm 0.5 \ \mathrm{km\,s^{-1}\,Mpc^{-1}}$, while the red vertical band represents the SH0ES measurement $H_0 = 73.5 \pm 1.4 \ \mathrm{km\,s^{-1}\,Mpc^{-1}}$.}\label{fig_h_evidence}
\end{figure}

In Fig.~\ref{fig_h_evidence}, we show the inferred values of the Hubble parameter $h$ for each cosmological model. It can be observed that none of the dark energy models alleviate the Hubble tension, as their inferred $h$ values remain below the \textit{Riess} predictions \cite{riess20113,riess2022comprehensive} but consistent with the \textit{Planck} predictions \cite{aghanim2020planck}. This can be understood in two ways. First, dynamical dark energy models cannot reduce the sound horizon since dark energy is sub-dominant at recombination. Therefore, if these models attempt to increase the inferred value of $h$, they become inconsistent with BAO measurements. Second, in all dynamical dark energy models predicting $w > -1$, the dark energy density scales as $f_{\mathrm{DE}} > 1$, leaving $\Omega_m h^2$ nearly unchanged from the $\Lambda$CDM prediction. To maintain consistency with the observed $H(z)$ data, a larger $f_{\mathrm{DE}}$ requires a lower inferred $H_0$, thereby intensifying the tension. Similar behavior has been reported in \cite{lee2022local,colgain2025much}, where the CPL model also favors $w_0 > -1$, leading to a reduced $H_0$ when DESI~DR2 is combined with CMB and SNe~Ia data.

\begin{figure*}
\begin{subfigure}{0.33\textwidth}
\includegraphics[width=\linewidth]{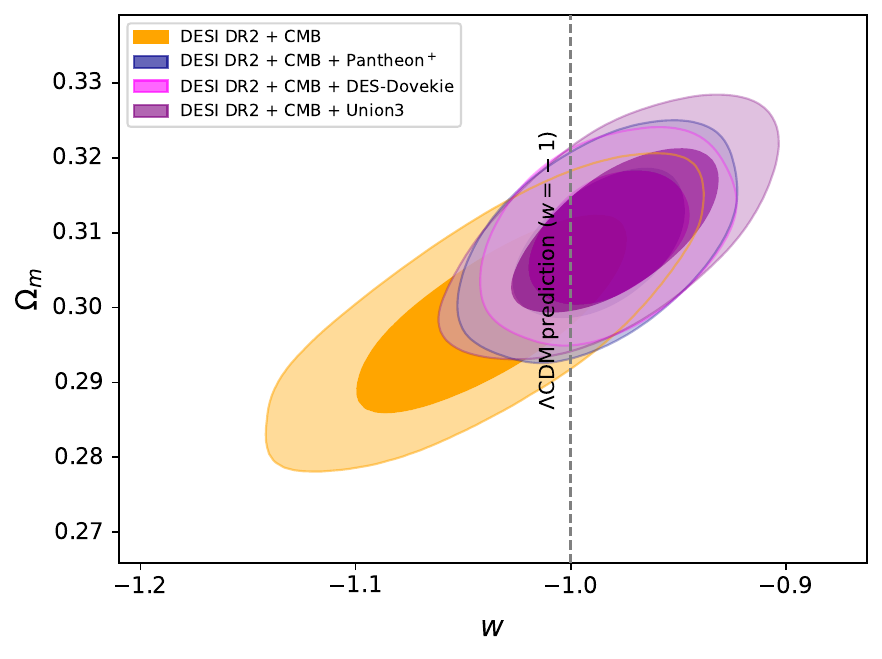}
    \caption{$w$CDM Parameterization}\label{fig_5a}
\end{subfigure}
\hfil
\begin{subfigure}{0.33\textwidth}
\includegraphics[width=\linewidth]{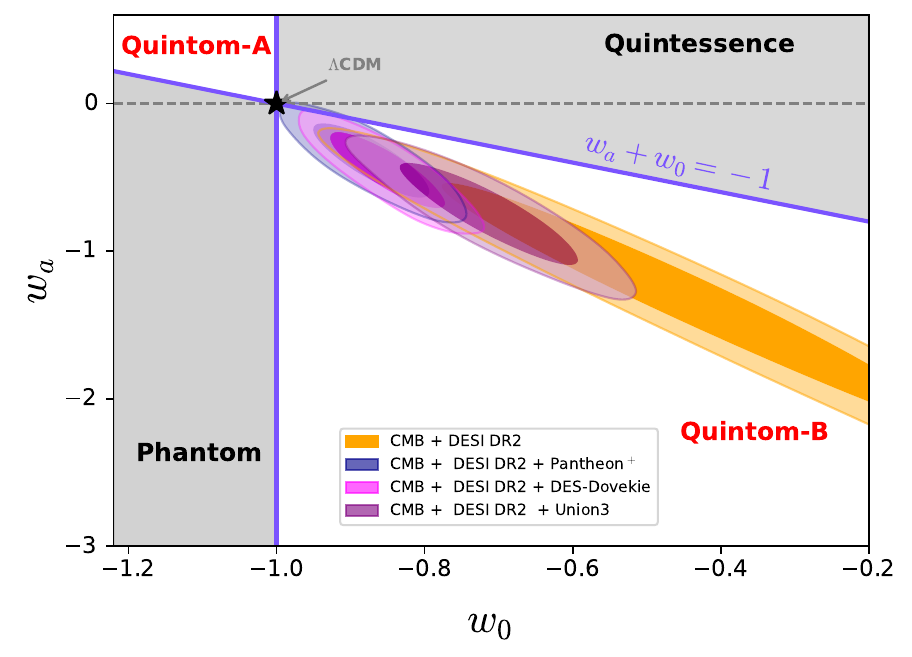}
    \caption{Logarithmic Parameterization}\label{fig_5b}
\end{subfigure}
\hfil
\begin{subfigure}{0.33\textwidth}
\includegraphics[width=\linewidth]{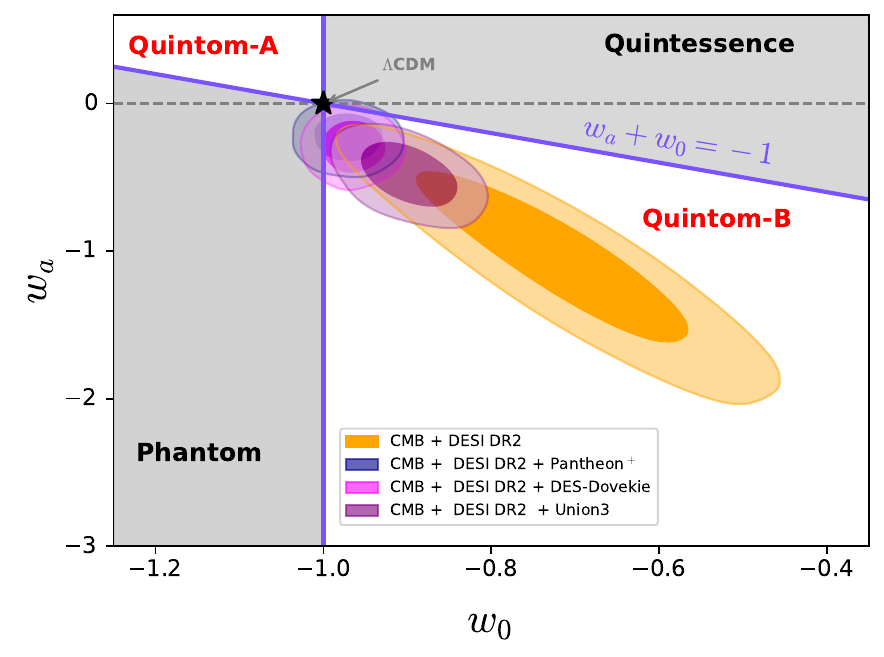}
    \caption{Exponential Parameterization}\label{fig_5c}
\end{subfigure}
\vskip\baselineskip
\begin{subfigure}{0.33\textwidth}
\includegraphics[width=\linewidth]{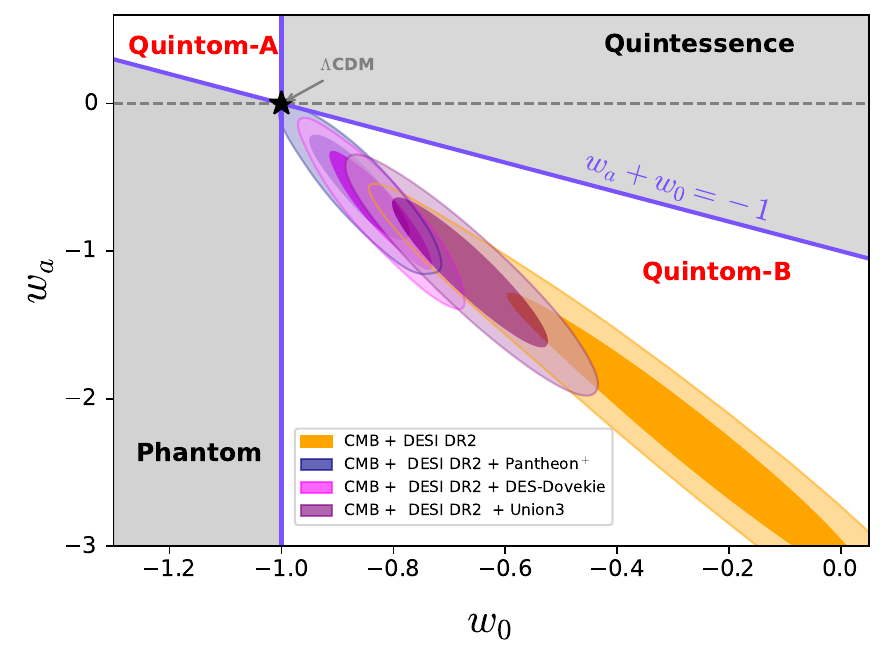}
    \caption{CPL Parameterization}\label{fig_5d}
\end{subfigure}
\hfil
\begin{subfigure}{0.33\textwidth}
\includegraphics[width=\linewidth]{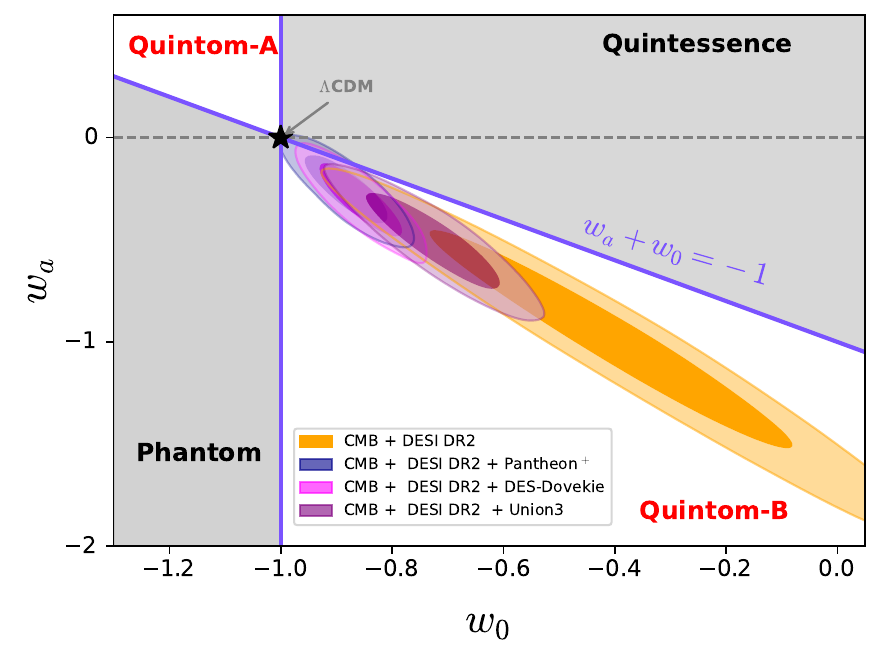}
     \caption{BA Parameterization}\label{fig_5e}
\end{subfigure}
\hfil
\begin{subfigure}{0.33\textwidth}
\includegraphics[width=\linewidth]{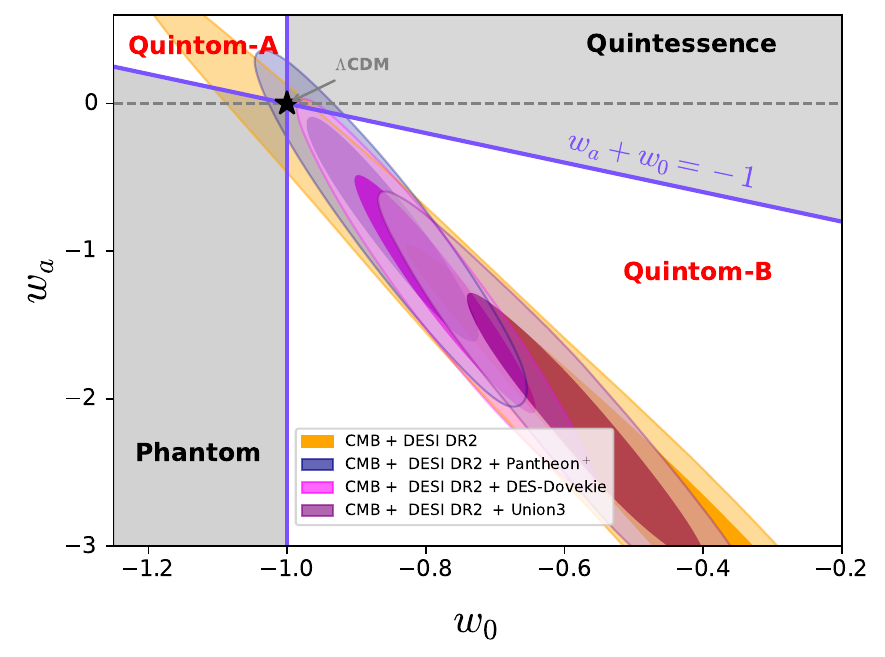}
     \caption{JBP Parameterization}\label{fig_5f}
\end{subfigure}
\begin{subfigure}{0.33\textwidth}
\includegraphics[width=\linewidth]{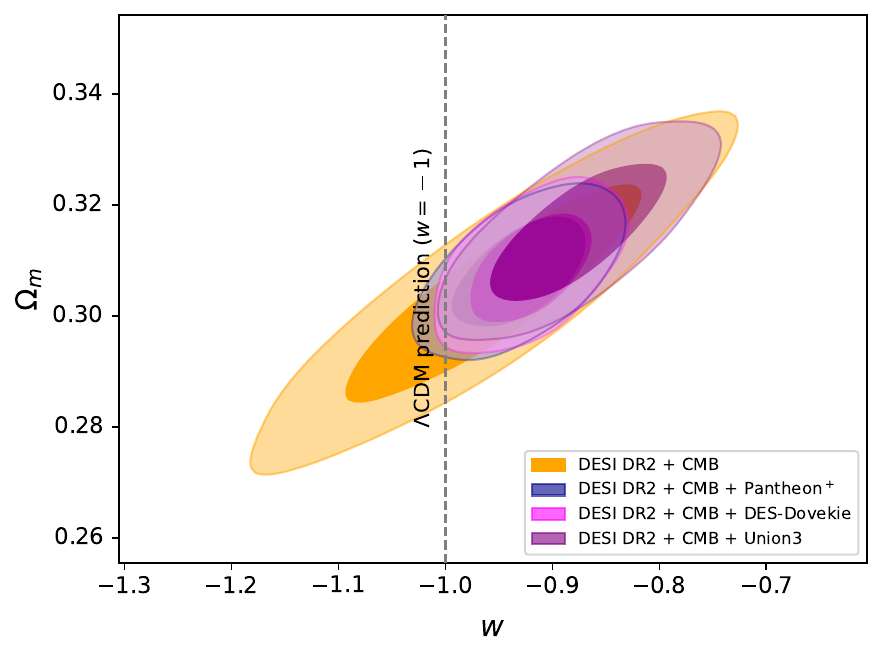}
    \caption{Thawing Parameterization}\label{fig_5g}
\end{subfigure}
\hfil
\begin{subfigure}{0.33\textwidth}
\includegraphics[width=\linewidth]{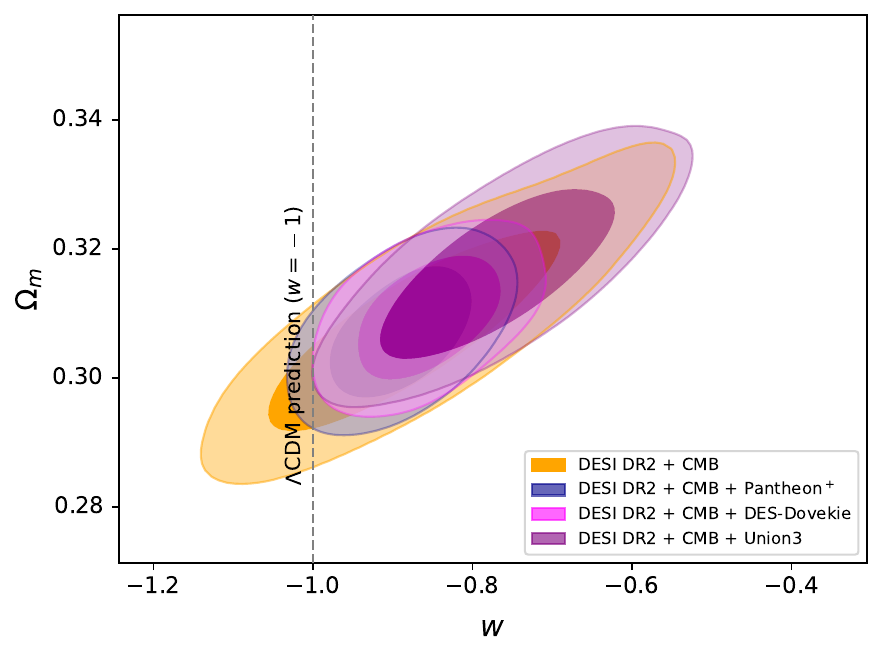}
     \caption{Mirage Parameterization}\label{fig_5h}
\end{subfigure}
\hfil
\begin{subfigure}{0.33\textwidth}
\includegraphics[width=\linewidth]{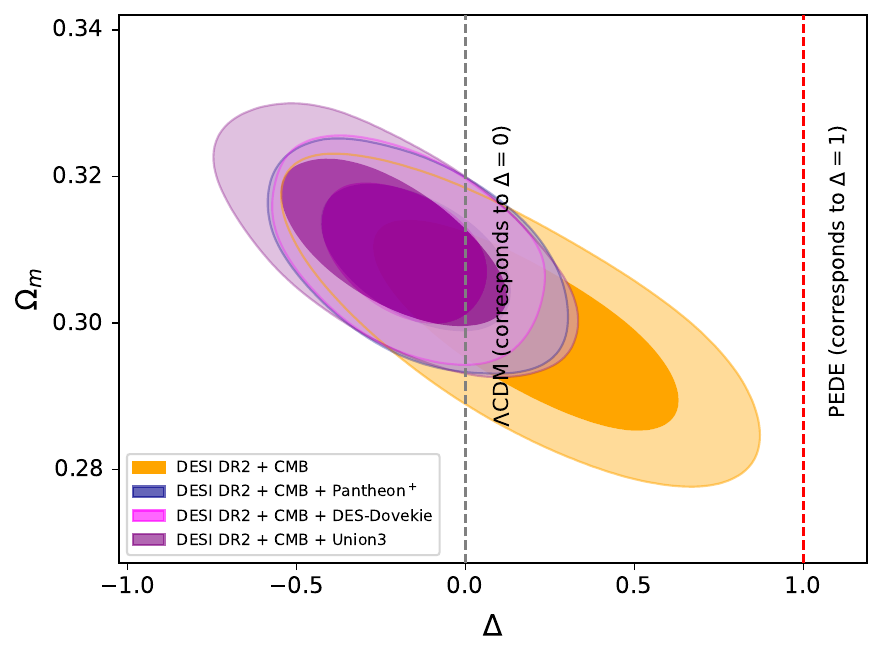}
     \caption{GEDE Parameterization}\label{fig_5i}
\end{subfigure}
\caption{The figure shows the posterior distributions of different planes of the $w$CDM, Logarithmic, Exponential, CPL, BA, JBP, Thawing, Mirage, and GEDE models using DESI DR2 measurements combined with different SNe Ia compilations (Pantheon$^+$, DES-Dovekie, Union3) and CMB compressed likelihood, at the 68\% (1$\sigma$) and 95\% (2$\sigma$) confidence intervals.}\label{fig_5}
\end{figure*}

Now we turn to the main observation of this paper. In Fig.~\ref{fig_5}, we present the relevant parameter planes. Specifically, Fig.~\ref{fig_5a} shows the $\Omega_m - w_0$ plane for the $w$CDM model. Our results, based on the combination of DESI DR2 and CMB data, are consistent with the findings of \cite{karim2025desi} ($w = -1.005 \pm 0.036$). Furthermore, our results also agree with the constraints obtained from DESI DR2 combined with CMB data and various SNe~Ia measurements, as reported by \cite{karim2025desi}: $w = -0.986 \pm 0.026$ with PP, $w = -0.983 \pm 0.024$ with DES-Dovekie, and $w = -0.979 \pm 0.031$ with Union3.

Figs.~\ref{fig_5b}, \ref{fig_5c}, \ref{fig_5d}, \ref{fig_5e}, and \ref{fig_5f} show the $w_0$-$w_a$ plane for the Logarithmic, Exponential, CPL, BA, and JBP redshift parameterizations, respectively. In each $(w_0,w_a)$ plane, the point $(w_0 =-1,w_a=0)$ corresponds to the $\Lambda$CDM model. It is worth noting that for each DE model, when combining the DESI DR2 measurements with CMB data and different Type~Ia supernova datasets, the inferred constraints favor the region $w_0 > -1$ and $w_a < 0$. This quadrant is also characteristic of the Quintom-B scenario, defined by $(w_0 > -1,\ w_a < 0,\ w_0 + w_a < -1)$, in which the equation of state evolves from phantom dark energy to quintessence consistent with previous findings in the literature \cite{cai2025quintom,ye2025hints,DESI:2025wyn}.

Figs.~\ref{fig_5g} and \ref{fig_5h} show the $w_0 - \Omega_m$ plane for the calibrated Thawing and Mirage redshift parameterizations. In the case of the calibrated Thawing model, we observe that for each combination of DESI DR2 with the different measurements of SNe Ia and CMB, the value of $w_0 > -1$ results in $w_a < 0$. Similarly, in the case of the calibrated Mirage model, combining DESI DR2 with CMB and each SNe Ia dataset yields results similar to those from the calibrated Thawing model, with $w_0 > -1$ in all cases and, consequently, $w_a < 0$. In Fig.~\ref{fig_5i}, which shows the $\Delta - \Omega_m$ plane for the GEDE model. It can be seen that the combination of DESI DR2 and CMB data yields $\Delta = 0.13_{-0.30}^{+0.26}$, consistent with the $\Lambda$CDM model. However, when different SNe Ia datasets such as Pantheon$^+$, DES-Dovekie, and Union3 are included, the GEDE model predicts a negative value of $\Delta$, indicating an injection of DE at high redshifts. A similar behavior can also be observed in previous studies \cite{lodha2025desi,lodha2025extended}.



\subsection{Statistical Analysis}
\begin{figure*}
\centering
\includegraphics[scale=0.6]{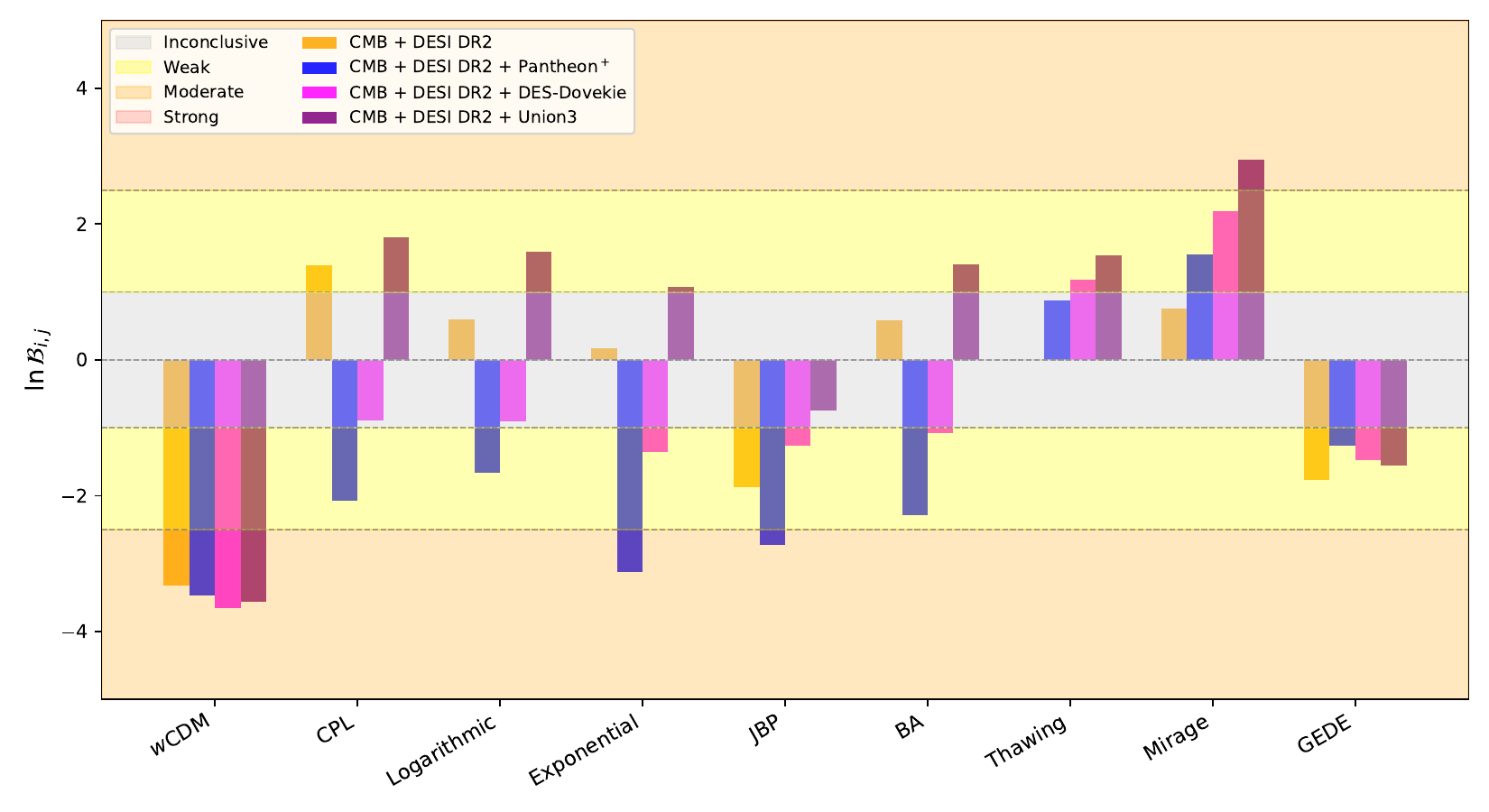}
\caption{The figure shows the relative difference ($\ln \mathcal{B}_{i,j}$) of the Bayes factor in logarithmic space compared to the $\Lambda$CDM model, using DESI DR2 measurements combined with different SNe Ia compilations (Pantheon$^+$, DES-Dovekie, Union3) and the CMB compressed likelihood.}\label{fig_bayes}
\end{figure*}

In Fig.~\ref{fig_bayes}, we show the comparative analysis of different cosmological models relative to the baseline $\Lambda$CDM model using the Bayes factor in logarithmic space, $\ln \mathcal{B}_{i,j}$. For the CMB + DESI DR2 combination, we find that the $w$CDM, JBP, and GEDE models show moderate to weak evidence in favour of $\Lambda$CDM, while the CPL, Logarithmic, Exponential, BA, and Mirage models show weak to inconclusive evidence relative to $\Lambda$CDM, with the Thawing model showing no clear preference. When Pantheon$^+$ data are added to the CMB + DESI DR2 combination, we find that the $w$CDM, CPL, Logarithmic, Exponential, JBP, BA, and GEDE models show weak to moderate evidence in favour of $\Lambda$CDM, whereas the Thawing and Mirage models show weak evidence relative to $\Lambda$CDM. Further, for the combination of DES-Dovekie with CMB + DESI DR2, we find that the $w$CDM, CPL, Logarithmic, Exponential, JBP, BA, and GEDE models show weak to moderate evidence in favour of $\Lambda$CDM, while the Thawing and Mirage models show weak to moderate evidence relative to $\Lambda$CDM. Finally, when Union3 is combined with CMB + DESI DR2, we find that the $w$CDM, JBP, and GEDE models show moderate to weak evidence in favour of $\Lambda$CDM, while the CPL, Logarithmic, Exponential, BA, Thawing, and Mirage models show weak to moderate evidence relative to $\Lambda$CDM.

We also quantify the preference for dynamical dark energy over the $\Lambda$CDM model across different data combinations, taking $\Lambda$CDM as the reference model. To assess this, we use the difference in the best-fit chi-square values between the DDE models and the $\Lambda$CDM scenario. Since $\Lambda$CDM is nested within the broader class of DDE models, $\Delta \chi^2_{\mathrm{MAP}}$ is expected to follow a $\chi^2$ distribution with $k$ degrees of freedom, corresponding to the number of additional parameters, assuming Gaussian uncertainties. The statistical significance is then expressed as an equivalent Gaussian $N\sigma$. $\mathrm{CDF}_{\chi^2}\left(\Delta \chi^2_{\mathrm{MAP}} \, | \, k~\mathrm{dof}\right)
= \frac{1}{\sqrt{2\pi}} \int_{-N}^{N} e^{-x^2/2} \, dx,$ where $k$ denotes the number of additional parameters in the DDE model relative to $\Lambda$CDM.

We find that $w$CDM shows no significant improvement over $\Lambda$CDM, with $N\sigma \approx 0$ for all dataset combinations, reaching at most $0.14\sigma$ for DES-Dovekie. For two-parameter models, CPL yields $N\sigma = 1.65,\,1.13,\,1.55,$ and $2.06$ for CMB+DESI, Pantheon$^+$, DES-Dovekie, and Union3, respectively. The Logarithmic parametrization gives $N\sigma = 1.59,\,1.34,\,1.65,$ and $2.12$, while the Exponential model gives $N\sigma = 1.64,\,1.19,\,1.59,$ and $2.08$. The JBP model shows lower values, $N\sigma = 1.10,\,0.83,\,1.38,$ and $1.82$. The BA model gives $N\sigma = 1.66,\,1.27,\,1.59,$ and $2.09$. For one-parameter extensions, the Thawing model yields $N\sigma = 1.65,\,1.96,\,2.10,$ and $2.22$, while the Mirage model gives $N\sigma = 1.73,\,1.98,\,2.24,$ and $2.32$, with the highest values obtained for Union3. The GEDE model shows smaller deviations, with $N\sigma = 0.80,\,0.97,\,1.11,$ and $1.07$ across the four dataset combinations.

Indeed, in the case of the CPL model, our results can be compared with previous DESI DR1 and DR2 analyses, which reported a preference for dynamical dark energy at the level of $2.6\sigma$, $2.5\sigma$, $3.5\sigma$, and $3.9\sigma$ when DESI DR1 is combined with CMB and Pantheon$^+$, Union3, and DESY5 datasets, respectively \cite{adame2025desi}. Similarly, in DESI DR2, the preference increases to $3.1\sigma$, $2.8\sigma$, $3.8\sigma$, and $4.2\sigma$ for the same dataset combinations \cite{karim2025desi}. In those analyses, the full CMB information is used from the temperature ($TT$), polarization ($EE$), and cross ($TE$) power spectra from \textit{Planck}, using the Commander likelihood for $\ell < 30$ and CamSpec for $\ell \geq 30$ \cite{Efstathiou2021Detailed,rosenberg2022cmb}, together with the combination of \textit{Planck} and ACT DR6 CMB lensing \cite{carron2022cmb,madhavacheril2024atacama,carron2022planck,qu2024atacama}.

\cite{Li:2025vuh} also reports evidence for dynamical dark energy in light of DESI DR2, combined with joint ACT \cite{louis2025atacama}, SPT \cite{SPT-3G:2026krg}, and \textit{Planck} data \cite{Efstathiou2021Detailed, rosenberg2022cmb}, for different dark energy parametrizations. In the case of the CPL, JBP, and BA models, previous studies combining DESI DR2 with joint ACT, SPT, and \textit{Planck} likelihoods report significantly higher deviations from $\Lambda$CDM. In particular, the CPL model shows $N\sigma = 2.3$-$3.9$, while the JBP model gives $N\sigma = 2.9$-$4.1$, and the BA model reaches $N\sigma = 3.0$-$4.2$, depending on the dataset combination. These values are consistently higher than those obtained in the present analysis, where we find $N\sigma \sim 1.1$-$2.1$ for the same class of models. In contrast, the present work focuses primarily on late-time measurements, combining DESI DR2 with SNe~Ia datasets and compressed CMB likelihoods, with particular emphasis on the preference for dynamical dark energy in the late-time datasets.

\subsection{Evolution of the EoS and the energy density function}
\begin{figure*}
\centering
\includegraphics[scale=0.28]{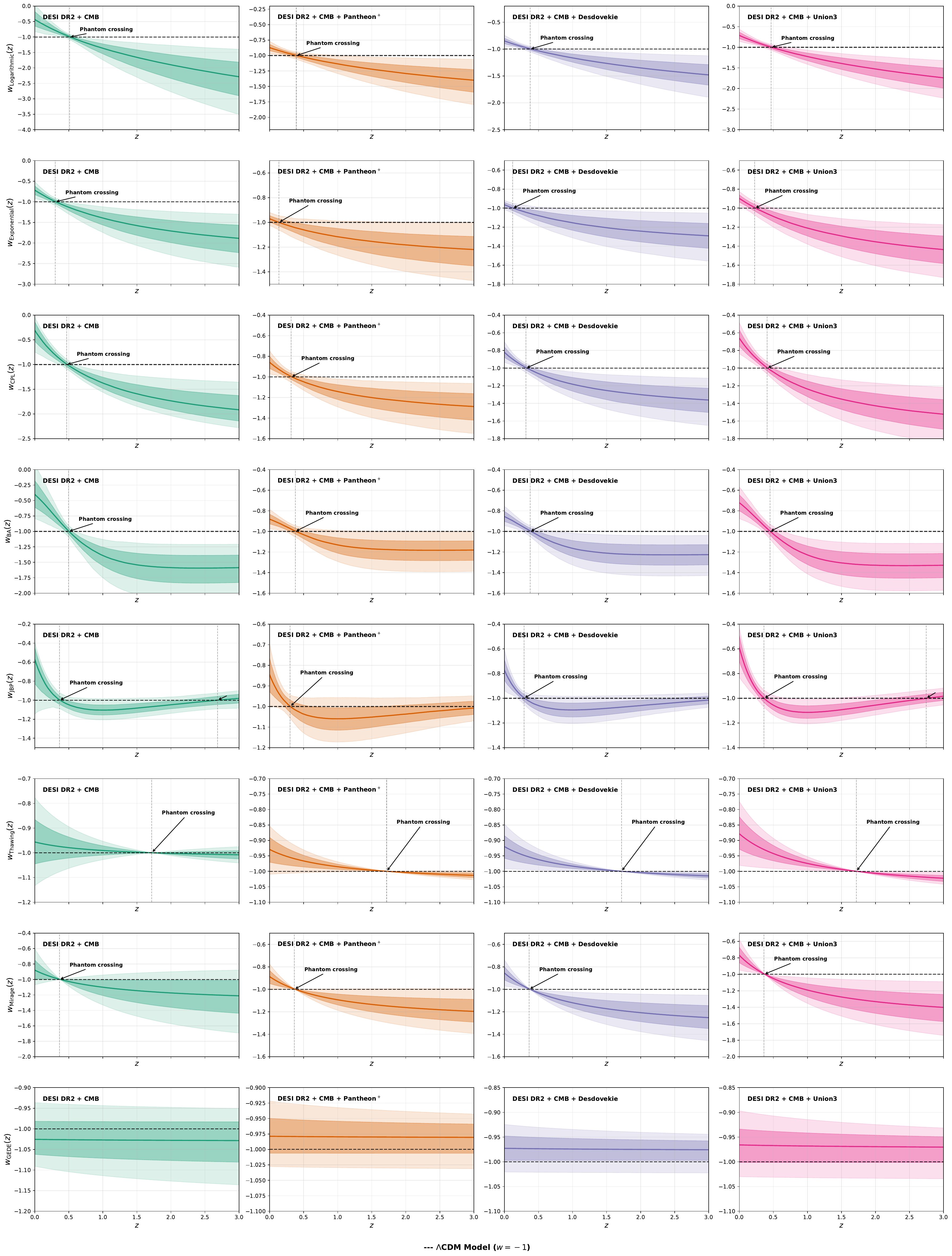}
\caption{This figure shows the evolution of $w(z)$ as a function of redshift, using DESI DR2 measurements combined with different SNe Ia compilations (Pantheon$^+$, DES-Dovekie, Union3) and CMB compressed likelihood}\label{fig_9}
\end{figure*}

\begin{figure*}
\centering
\includegraphics[scale=0.28]{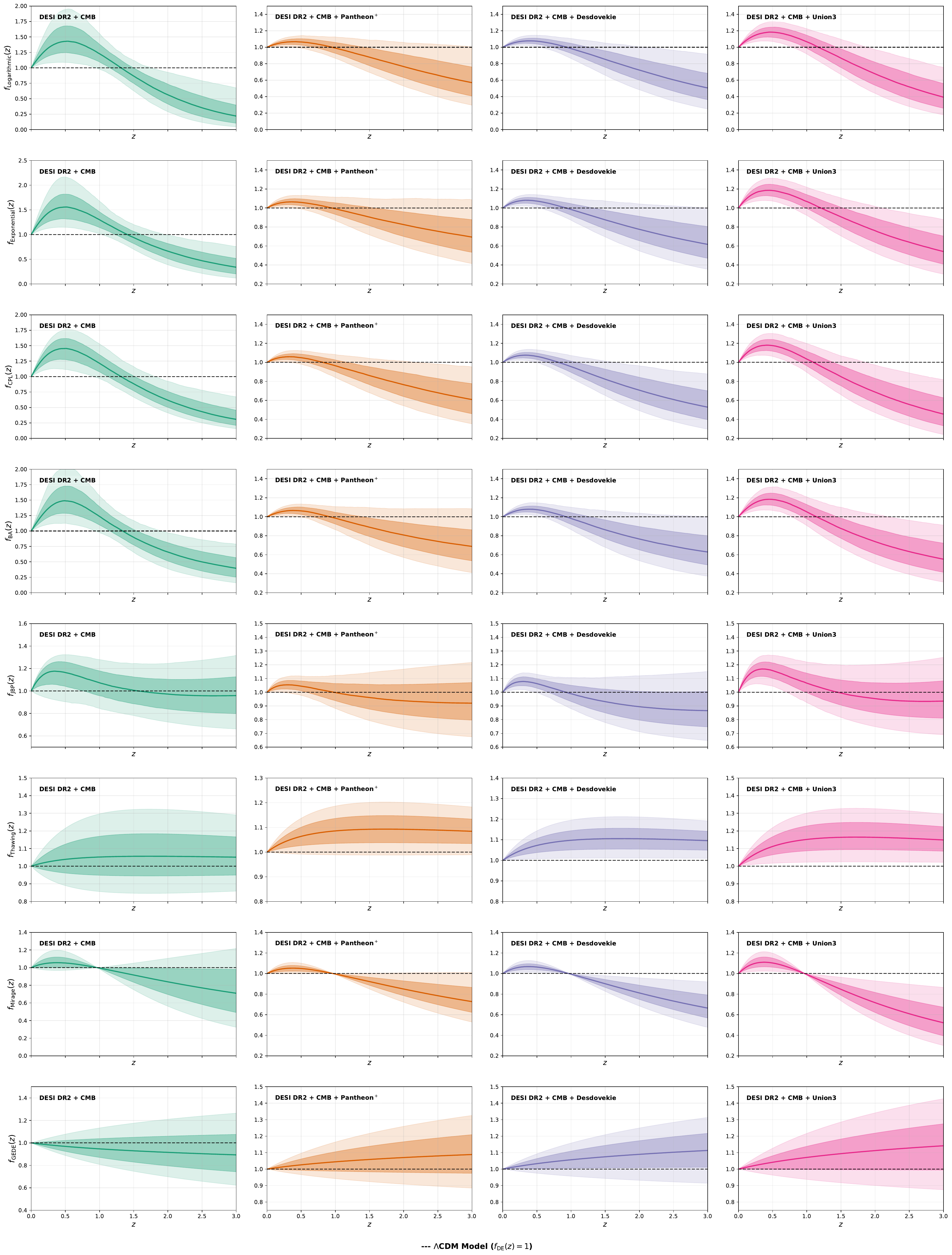}
\caption{This figure shows the evolution of $f_{DE}(z)$ as a function of redshift, using DESI DR2 measurements combined with different SNe Ia compilations (Pantheon$^+$, DES-Dovekie, Union3) and CMB compressed likelihood}\label{fig_10}
\end{figure*}

In this subsection, we show the redshift evolution of the EoS parameter in Fig.~\ref{fig_9} and the energy density function in Fig.~\ref{fig_10}, in order to gain a deeper physical understanding of the models. Fig.~\ref{fig_9} shows that, for the Logarithmic, Exponential, CPL, BA, JBP, and Mirage models, and for all dataset combinations, $w(z)$ falls below the cosmological constant boundary $w = -1$ at redshifts $z \gtrsim 0.5$, indicating that each model predicts values in the phantom regime ($w < -1$). At lower redshifts, $z \lesssim 0.5$, $w(z)$ rises again above $w = -1$, predicting values in the quintessence-like regime ($w > -1$). This crossing, where $w(z)$ rises from the phantom regime at high redshift to the quintessence-like regime at low redshift by crossing the boundary $w = -1$, is called a \textit{phantom crossing.} This evidence of phantom crossing in our analysis is consistent with previous studies reported in the literature \cite{ozulker2025dark,chen2025quintessential,silva2025testing,roy2025phantom,guedezounme2025phantom,hu2005crossing,yao2025general}. This kind of behavior is characterized by the Quintom-B–type dark energy scenario, and is consistent with previous findings in the literature \cite{cai2025quintom,ye2025hints,DESI:2025wyn}. Indeed, it is interesting to note that the JBP model crosses the $w = -1$ boundary twice, at $z \gtrsim 0.5$ as well as in the redshift range $2.0 \lesssim z \lesssim 3.0$.

While in the case of the Thawing model this crossing occurs in the redshift range $1.5 \lesssim z \lesssim 2.0$. Finally, in the GEDE model, for the DESI DR2 + CMB dataset combination, the model predicts values in the phantom regime during the entire redshift evolution. In contrast, for the DESI DR2 + CMB + Pantheon$^+$, DESI DR2 + CMB + DES-Dovekie, and DESI DR2 + CMB + Union3 datasets, the predicted values remain above $w = -1$ throughout the full evolution, indicating that there is no phantom crossing in the GEDE model for these dataset combinations.

We find that the combination of DESI DR2 with different SNe~Ia samples and compressed CMB likelihoods exhibits a quintom-like evolution of dark energy. In this context, it is important to revisit the \textit{No-Go theorem} associated with dynamical dark energy \cite{Cai:2009zp,Qiu:2010ux}. In particular, a single canonical scalar field or a single perfect fluid model cannot reproduce such quintom-like behavior, as these frameworks forbid the equation-of-state parameter $w$ from crossing the cosmological constant boundary $w = -1$ \cite{Vikman:2004dc,Deffayet:2010qz}.

It is important to emphasize that such behavior can also be partly induced by the choice of the EoS parametrization, and should therefore be interpreted with caution. In particular, parameters such as $(w_0, w_a)$ correspond to an effective parametrization fitted over a limited redshift range, and extrapolating this parametrization over the full cosmic history can be misleading. While it is often stated that quintessence models lie above the line $w_0 + w_a = -1$, this is not strictly correct. Although the true equation of state $w(a)$ for canonical quintessence cannot cross the phantom divide $w = -1$, the fitted parametrization may still occupy regions in the $(w_0, w_a)$ plane that formally imply a phantom crossing when extrapolated beyond the fitted redshift range.

Recent studies have demonstrated that quintessence models can populate more negative regions of the $(w_0, w_a)$ plane when the parametrization is fitted to the model \cite{Wolf:2023uno,shlivko2024assessing,Wolf:2024eph}, and have also highlighted subtleties in the interpretation of such parametrizations \cite{Cortes:2024lgw}. Therefore, an apparent phantom crossing inferred from $(w_0, w_a)$ should not be directly interpreted as evidence for true phantom dynamics, but rather as a feature of the effective parametrization within the redshift range probed by the data.

Fig.~\ref{fig_10} shows the redshift evolution of the energy density function for the CPL, Logarithmic, Exponential, JBP, BA, and Mirage models. For all dataset combinations, $f_{DE}(z)$ crosses $f_{DE}=1$ and converges to unity at the present epoch, $z=0$, i.e., $f_{DE}(0)=1$. In contrast, for the Thawing and GEDE models, $f_{DE}(z)$ remains above unity ($f_{DE}>1$) throughout the evolution and only approaches $f_{DE}(0)=1$ at $z=0$. In both figures, the solid lines represent the mean values, while the light and dark shaded regions correspond to the $1\sigma$ and $2\sigma$ confidence intervals, respectively.


\section{Conclusions}\label{sec_5}
This study provides evidence suggestive of the evolving nature of dark energy, though not yet conclusive, challenging the core assumptions of the standard $\Lambda$CDM model with a cosmological constant. In particular, our analysis of DESI DR2 indicates that the LRG1 and LRG3+ELG1 tracers introduce tensions in the inferred value of $\Omega_m$, highlighting their key role in driving deviations from $\Lambda$CDM.

Extending the analysis beyond $\Lambda$CDM, we find that the inclusion of LRG1-2 tracers systematically shifts the inferred parameters of dynamical dark energy models, particularly toward $w_0 > -1$. When combined with CMB and SNe~Ia datasets, the constraints in the $w_0$-$w_a$ plane tend to occupy the region $(w_0 > -1,\, w_a < 0)$, suggesting a quintom-like evolution at a phenomenological level.

Using the $\Delta\chi^2_{\mathrm{MAP}}$ statistic, we find that the preference for dynamical dark energy remains at the weak-to-moderate level, typically corresponding to $N\sigma \sim 1$-$2.3$, depending on the model and dataset combination, with the largest values generally obtained for the Union3 dataset. Thus, no dynamical dark energy model is consistently preferred over $\Lambda$CDM, and current data provide only mild, non-decisive hints of evolution.

The reconstructed evolution of the equation-of-state parameter $w(z)$ exhibits indications of phantom crossing behavior in several parametrizations. However, according to the theoretical no-go theorem, such behavior cannot arise in single canonical scalar-field or perfect fluid models, and should therefore be interpreted with caution. In addition, part of this behavior may be induced by the choice of parametrization, rather than reflecting a fundamental physical effect.

The Bayesian analysis further supports this interpretation, showing that $\Lambda$CDM remains statistically competitive across all dataset combinations, with no model demonstrating a robust or consistent preference over it. These results indicate that, while phenomenological features of evolving dark energy appear in certain cases, they do not yet provide decisive evidence for a departure from the concordance cosmological model.

These findings highlight that our current understanding of DE may be incomplete, although present data do not yet require a departure from the $\Lambda$CDM framework. If confirmed, the evolving nature of DE could significantly impact our understanding of the Universe expansion and may require a reevaluation of the fundamental assumptions underlying cosmology. Therefore, further investigation and more precise measurements, particularly at low redshifts, are essential to fully characterize the nature of DE and its role in shaping the cosmos.

\section*{Acknowledgements}
SC acknowledges the Istituto Nazionale di Fisica Nucleare (INFN) Sez. di Napoli,  Iniziative Specifiche QGSKY and MoonLight-2  and the Istituto Nazionale di Alta Matematica (INdAM), gruppo GNFM, for the support. This paper is based upon work from COST Action CA21136 -- Addressing observational tensions in cosmology with systematics and fundamental physics (CosmoVerse), supported by COST (European Cooperation in Science and Technology).

\bibliographystyle{elsarticle-num}
\bibliography{mybib}

\end{document}